\documentclass[12pt]{article}

\usepackage{amssymb}
\usepackage[mathscr]{euscript}
\usepackage{graphicx}

\newtheorem{prop}{Proposition}[section]

\newtheorem{lemm}{Lemma}[section]
\newtheorem{theo}{Theorem}[section]
\newtheorem{coro}{Corollary}[section]

\newcommand{\iN}{\hbox{ {\leaders\hrule\hskip.2cm}{\vrule height .22cm} }}

\newcommand{\R}[1][]{\ensuremath{{\mathbb{R}^{#1}} }}
\newcommand{\C}[1][]{\ensuremath{{\mathbb{C}^{#1}} }}

\newcommand{\z}{\textsf{z}}

\newcommand{\y}{\textsf{y}}
\newcommand{\ypoint}{\dot{\textsf{y}}}

\newcommand{\etazero}{\mathop{\eta}\limits^{\oldstylenums{0}}}
\newcommand{\etaun}{\mathop{\eta}\limits^{\oldstylenums{1}}}

\usepackage{color}
\definecolor{warmblack}{rgb}{0.0, 0.26, 0.26}
\definecolor{slategray}{rgb}{0.44, 0.5, 0.56}
\definecolor{darkjunglegreen}{rgb}{0.1, 0.14, 0.13}
\usepackage[colorlinks,linkcolor=slategray,citecolor=darkjunglegreen,urlcolor=darkjunglegreen]{hyperref}
  
\title{Curved space-times by crystallization\\of liquid fiber bundles}
\author{Fr\'ed\'eric H\'ELEIN\footnote{
IMJ-PRG, UMR CNRS 7586, Universit\'{e} Paris 7--Paris Diderot, UFR de Math\'{e}matiques,  Case 7012,
B\^{a}timent Sophie Germain, 75205 Paris Cedex 13, France; \textsf{helein@math.jussieu.fr}},
Dimitri VEY\footnote{Nomad Institute for Quantum Gravity, France;
\textsf{dim.vey@gmail.com}} }

\begin{document}
\maketitle

\emph{Abstract} --- Motivated by the search for a Hamiltonian formulation of
Einstein equations of gravity which depends in a minimal way on choices of coordinates, nor
on a choice of gauge, we develop a
multisymplectic formulation on the total space of the principal
bundle of orthonormal frames on the 4-dimensional space-time. This leads quite naturally
to a new theory which takes place on 10-dimensional manifolds. The fields are
pairs of $((\alpha,\omega),\varpi)$, where $(\alpha,\omega)$ is a 1-form with coefficients in
the Lie algebra of the Poincar{\'e} group and $\varpi$ is an 8-form with coefficients in the dual
of this Lie algebra. The dynamical equations derive from a simple variational principle and imply
that the 10-dimensional manifold looks locally like the total space of a fiber bundle over a
4-dimensional base manifold. Moreover this base manifold
inherits a metric and a connection which are solutions of a system of Einstein--Cartan equations.

\section{Introduction}

A well-known fundamental feature of General Relativity is the fact that the shape of space-time and the structures which
determine its physical properties (i.e. the metric, the connection) are not fixed a priori, but by the dynamics.
However if we lift the description of this theory to the principal bundle of orthonormal moving frames
over the space-time, we need to assume a priori constraints on its total space, namely the axioms of the definition
of a principal bundle and of a connection.

Our approach in this paper leads to release the 10-dimensional total space of the principal bundle and
the connection from these a priori constraints. We will start from a 10-dimensional manifold (the dimension
of the Poincar{\'e} group) which can be considered as a white sheet: we don't draw on it the fibers
of the principal bundle, nor a fortiori the way to quotient out this manifold to get a 4-dimensional space-time.
We will instead recover these structures from the dynamical equations. Eventually we also derive from the
dynamical equations the existence of a metric and connection on a 4-dimensional quotient manifold
which satisfy some Einstein--Cartan system of equations. The resulting theory has hence some
flavor of a Kaluza--Klein theory.

The way we achieve our theory is not based on some ad hoc construction, but on a study of the
Hamiltonian structure of Einstein equations, starting from the variational Weyl--Einstein--Cartan formulation
(called WEC in this paper and erroneously known as the Palatini one, see \cite{ferraris}).
This analysis is guided by a philosophical standpoint: to privilege formulations
which are as covariant as possible, which means mathematically that we look for a formulation which
depends in a minimal way on choices of coordinates. While several alternative theories exist
for that purpose, we favour here the multisymplectic approach, since it simultaneously
respects in a natural way the locality of physical theories. In a few words (see also below)
the basic idea of the multisymplectic
formalism, which goes back to V. Volterra, is to consider all first order derivatives of the fields
as analogues of the velocity in Mechanics and to perform the Legendre transform with respect to
all these first order derivatives.

But for a gauge theory, as for instance the WEC formulation of gravity which involves moving
frames, this is not enough, since in the standard description of gauge theories
the coordinates independence of the multisymplectic
formalism is spoiled by the need for choosing a particular gauge
(here a particular moving frame) for writing the equations.
In order to go beyond these difficulties we follow the approach\footnote{An alternative approach 
would consist in building a suitable reduction of
the geometry of connections on a $\mathfrak{G}$-principal bundle as for instance
in \cite{Bruno05a,Bruno05b}.} in \cite{helein14} which consists in lifting
the problem on the total space of the principal bundle involved, as pictured by C. Ehresmann.
We hence will meet and use ideas and points of view developped by E. Cartan \cite{Cartan23,Cartan35},
including his theory of the equivalence
problem.\\

\noindent
\emph{Aknowledgements:} we thank Friedrich W. Hehl and Igor Kanatchikov for comments on a first version of this
paper.

\subsection{Overview of the paper}

The origin of the multisymplectic formalism goes back to the discovery by Volterra at the end of
the ninetieth century 
\cite{volterra1,volterra2} of generalizations
of the Hamilton equations for variational problems with several variables. These ideas were first
developped in particular by C. Carath{\'e}odory  \cite{Caratheodory29}, T. De Donder \cite{Donder01}, H. 
Weyl \cite{Weyl35},  T.  Lepage \cite{Lepage}, and later by  P.  Dedecker \cite{Dedecker53,Dedecker77}. 
In the seventies  of the past century, this  theory was
geometrized in a way analogous to the construction of symplectic geometry by several mathematical
physicists. In particular,  the Polish school  formulated important ideas and   developed the
{\it multiphase-space} formalism in the work of  W.M. Tulczyjew \cite{Tulczyjew68}, J.   Kijowski \cite{KS0},
 Kijowski and Tulczyjew \cite{JKWMT}, Kijowski    and  W. Szczyrba \cite{ KS1,KS2}. Parallel to
this development, the paper  by H.  Goldschmidt and  S. Sternberg   \cite{HGSS}  gave a formulation
of the Hamilton equations in terms of  the Poincar\'e-Cartan form and  the underlying jet
bundles geometry,   and a related approach was also developed by   the Spanish school in
P.L. Garc\'{i}a \cite{Garcia00}, Garc\'{i}a and A. P\'{e}rez-Rend\'{o}n  \cite{Garcia02}.
This theory has many recent developments
which we cannot report here (see e.g.
\cite{Cantrijn99,Leon14,forgergomes,forgerromero,GIMMSY,Gotay91,hk1,hk2,kanatchikov1,lopez-marsden}).
Today the Hamilton--Volterra equations are often
called the De Donder--Weyl equations for reference to \cite{Donder01,Weyl35}, which is inaccurate
\cite{heleinleeds}. 
In this paper we name them the {HVDW} equations for Hamilton--Volterra--De Donder--Weyl.

A multisymplectic manifold is a smooth manifold $\mathcal{N}$ endowed with a \emph{multisymplectic} 
$(m+1)$-form $\omega$, i.e. $\omega$ is closed and one often assumes that it is non degenerate,
i.e. that the only vector field $\xi$ on the manifold such that $\xi\iN  \omega = 0$ is zero.
Here $m$ refers to the number of independent variables of the associated variational problem.
An extra ingredient is a Hamiltonian function $H:\mathcal{N}\longrightarrow \Bbb{R}$. One can then
describe the solutions of the {HVDW} equations by oriented $m$-dimensional
submanifolds $\Gamma$ of $\mathcal{N}$ which satisfy the condition that, at any point
$\textsc{m}\in \mathcal{N}$, there exists a basis $(X_1,\cdots,X_m)$ of $T_\textsc{m}\Gamma$
such that $X_1\wedge \cdots \wedge X_m\iN  \omega  = (-1)^mdH$.
Equivalently one can replace $\omega $ by its restriction to the level set $H^{-1}(0)$
and describe the solutions as the submanifolds $\Gamma$ of $H^{-1}(0)$ such that
$X_1\wedge \cdots \wedge X_m\iN \omega = 0$ everywhere (plus some independence conditions,
see e.g. \cite{heleinleeds}).

Before applying the multisymplectic formalism and in order to describe the theory in a way which does not depend on any 
choice of gauge, in Paragraph \ref{paragraphlifting} we first translate
and lift the 4-dimensional WEC variational principle
to the total space $\mathcal{P}$ of the principal bundle.
Note that this kind of approach shares some similarities with the use of Cartan geometries as
e.g. in \cite{Wise-01}.
Denoting by $\mathfrak{p}$ the Lie algebra of the Poincar{\'e} group,
the connection and the vierbein are both represented by a $\mathfrak{p}$-valued 1-form $\eta$ on
$\mathcal{P}$ which satisfies \emph{normalization} (\ref{normalisationStandard}),
(\ref{normalisationSoudureStandard})
and \emph{equivariance} (\ref{equivarianceStandard}), (\ref{equivarianceSoudureStandard}) hypotheses. Although a priori
necessary the equivariance condition has the drawback of being a \emph{non holonomic} constraint, {i.e.} on
the first order derivatives of the field, which, to our opinion, is not a natural condition.
Another preliminary step is, in Paragraph \ref{paragraphforget}, to forget the normalization condition and to express the equivariance
condition in a way which is independent on it,
but relies on Cartan's theory of the equivalence problem. The subsequent
computations will confirm that the normalization condition, as its name suggests it, is not
essential and can be recovered by a suitable choice of coordinates.

Then, in Section \ref{sectionmultipatates}, we apply the multisymplectic machinery for $m=10$ and compute the Legendre transform 
by treating connections as equivariant $\mathfrak{p}$-valued 1-forms on $\mathcal{P}$.
We find that the natural multisymplectic manifold can be built from the vector
bundles $\mathfrak{p}\otimes T^*\mathcal{P}$ and $\mathfrak{p}^*\otimes \Lambda^8T^*\mathcal{P}$
over $\mathcal{P}$, where $\mathfrak{p}^*$ its dual vector space of $\mathfrak{p}$. These vector
bundles are endowed with a canonical $\mathfrak{p}$-valued 1-form $\eta
= (\eta^0, \cdots, \eta^9)$ and a canonical
$\mathfrak{p}^*$-valued $8$-form $\psi = (\psi_0,\cdots,\psi_9)$ respectively. Then the multisymplectic
manifold is the submanifold $\mathcal{M}$ of the total space of the vector bundle
$(\mathfrak{p}\otimes T^*\mathcal{P}) \oplus_\mathcal{P}(\mathfrak{p}^*\otimes
\Lambda^8T^*\mathcal{P})$, defined by the equations
$\eta^a\wedge \eta^b \wedge \psi_A = \kappa^{ab}_A \eta^0\wedge \cdots\wedge \eta^9$,
$\forall a,b,A$ s.t. $0\leq a,b\leq 3$ and $0\leq A\leq 9$, where the coefficients
$\kappa^{ab}_A$ are some fixed structure constants. The manifold $\mathcal{M}$ is equipped
with the $10$-form $\theta = \psi \wedge  (d\eta + \eta \wedge \eta)$,
where the duality pairing between $\mathfrak{p}^*$ and $\mathfrak{p}$ is implicitly
assumed. The solutions of the Hamilton equations are sections $\varphi$ of $\mathcal{M}$ over $\mathcal{P}$
which are critical points of the action $\mathcal{A}[\varphi]=\int_\mathcal{P}\varphi^*\theta$.
At this stage we will decide to \emph{remove} the unnatural equivariance
constraints (on $\varphi^*\eta$) and we derive the corresponding generalized Hamilton equations
in Section \ref{sectionHamilton}. We note that the resulting theory is manifestly a gauge theory with gauge group 
the Poincar\'e group, whose importance for gravity theories is stressed in
\cite{hehl}.

Then several interesting phenomena occur. The first one is that the dynamical equations
force the manifold $\mathcal{P}$ to be locally fibered over a 4-dimensional manifold, with
6-dimensional fibers. This is the content of Lemma \ref{spontaneousfibration} in Paragraph \ref{paragraphfibration} (which follows
from similar mechanisms as in \cite{helein1}, see Lemma 2.1):
a metric and a connection emerge spontaneously from the solution on the 
4-dimensional quotient space. Moreover we can recover the normalization
conditions by a suitable choice of coordinates adapted to this local fibration and,
as in \cite{helein14} for the Yang--Mills fields, the dynamical equations force the
fields to satisfy the equivariance conditions along these fibers. The second phenomenon
appears after a long computation in Paragraph \ref{paragraphlong}, done in order to write the equations in coordinates adapted to 
these fibration. The metric and the connection on the 
4-dimensional quotient space  satisfy an Einstein--Cartan system of
equations (\ref{firstsystem})
\begin{equation}\label{ECsystem}
 \left\{
 \begin{array}{lcl}
  \hbox{E}{^b}_a & = & \frac{1}{2} \rho_j\cdot p{_a}{^{bj}}\\
   T{^a}_{cd} & = & - \left( \textsf{h}_{de}\delta^a_{a'}\delta^{c'}_c 
 + \frac{1}{2}\delta^{c'}_{a'}(\delta^a_d\textsf{h}_{ce}
 - \delta^a_c\textsf{h}_{de})\right) \rho_j\cdot p{_{c'}}{^{ea'j}}
 \end{array}
 \right.,
\end{equation}
where $\hbox{E}{^b}_a$ is the Einstein tensor, $T{^a}_{cd}$ is the torsion tensor,
$(\rho_j)_{4\leq j\leq 9}$ is a left invariant moving frame on the 6-dimensional fiber and
$\rho_j\cdot f$ is the derivative of $f$ with respect to $\rho_j$.
The right hand sides of
(\ref{ECsystem}) are covariant divergences involving derivatives with respect to coordinates on the fibers
of the tensors $p{_a}{^{bj}}$ and $p{_{c}}{^{eaj}}$, which are components
of $\varphi^*\psi$.
They play here the role of a stress-energy tensor
and an angular momentum tensor, respectively. The tensors $p{_a}{^{bj}}$ and $p{_{c}}{^{eaj}}$
satisfy also non homogeneous Maxwell type equations
(\ref{secondsystem}) which involve space-time partial derivatives 
and are defined up to some gauge transformations (see Section \ref{sectioninvariance}).

At this point come some difficulties but also some exciting and challenging questions,
discussed in Section \ref{sectiondiscussion}.
A naive wish would be that the r.h.s. of (\ref{ECsystem}) vanish, in order to recover the
standard vacuum Einstein equations of gravity. This is actually the case if we replace the Lorentz group by $SO(4)$
(or its universal cover $Spin(4)$):
then, as shown in Theorem \ref{theoremEuclide}, under reasonable hypotheses, one can show that
the r.h.s. of (\ref{ECsystem}) vanish and hence we recover exactly all the
orthonormal frame bundles of Einstein manifolds. The main reason here is that $SO(4)$ or $Spin(4)$
are \emph{compact}, as in \cite{helein14} for Yang--Mills. But $SO(1,3)$ is not compact and we cannot
conclude that the r.h.s. of (\ref{ECsystem}) vanish in general.
Hence we are led to consider a larger class of solutions than the classical Einstein metrics
in vacuum. One needs for that purpose to understand Equations
(\ref{secondsystem}) and to know whether one could assume natural and physically relevant hypotheses
on $p{_a}{^{bj}}$ and $p{_{c}}{^{eaj}}$ which would imply that the r.h.s. of
(\ref{ECsystem}) vanish or, at least, satisfy some equations (besides the usual conservation
law satisfied by the stress-energy tensor and the angular momentum tensor). It would be
also interesting to see whether the r.h.s. of
(\ref{ECsystem}) could be interpreted as a dark matter and/or a dark energy source.
In a broader framework, it would interesting to study similar models coupled with matter
fields and to understand the possible role of the extra fields $\varphi^*\psi$ (or their generalizations)
in the interaction between gravity and the other fields.

\subsection{Summary of notations}\label{notations}
\begin{itemize}
 \item $\mathbb{M}$ is a 4-dimensional real affine space and $\vec{\mathbb{M}}$ is
the associate vector space endowed with a non degenerate symmetric bilinear form $\textsf{h}$
(either the Minkowski metric or the standard Euclidean one);
$(E_0,E_1,E_2,E_3)$ is an orthonormal basis of $(\vec{\mathbb{M}},\textsf{h})$.
\item $\mathfrak{G}$ is the group of linear isometries of $(\vec{\mathbb{M}},\textsf{h})$
or its universal cover (either $SO(1,3)$ or $SL(2,\C)$ if
$\textsf{h}$ is the Minkowski metric or $SO(4)$ or $Spin(4)$ if $\textsf{h}$ is the Euclidean metric);
$\mathfrak{g}$ is the Lie algebra of $\mathfrak{G}$.
\item $(u_4,u_5,u_6,u_7,u_8,u_9)$ is a basis of $\mathfrak{g}$ and $c^k_{ij}$
($4\leq i,j,k\cdots \leq 9$) are the structure coefficients
of $\mathfrak{g}$ in this basis, so that $[u_i,u_j] = c^k_{ij}u_k$.
\item If $\mathcal{R}:\mathfrak{G}\longrightarrow GL(\vec{\mathbb{M}})$ is the standard linear representation,
then $\forall g\in \mathfrak{G}$,  $(g{^a}_b)_{0\leq a,b\leq 3}$ are the coefficients of the matrix of
$\mathcal{R}(g)$ in the basis $(E_0,E_1,E_2,E_3)$, i.e. $\mathcal{R}(g)(E_b) = E_ag{^a}_b$.
\item Similarly, if $\mathcal{R}:\mathfrak{g}\longrightarrow gl(\vec{\mathbb{M}})$
is the standard linear representation, $\forall \xi\in \mathfrak{g}$, 
$(\xi{^a}_b)_{0\leq a,b\leq 3}$ are the coefficients 
of the matrix of $\mathcal{R}(\xi)$ in the basis $(E_0,E_1,E_2,E_3)$. We then have $\xi^{ab}+\xi^{ba}=0$, where
$\xi^{ab} = \xi{^a}_{b'}\textsf{h}^{b'b}$, for $0\leq a,b,c,d,\cdots\leq 3$.
\item In particular, for $4\leq i\leq 9$, $(u^a_{ib})_{0\leq a,b\leq 3}$ are the coefficients 
of the matrix of $\mathcal{R}(u_i)$; we set $u^{ab}_i:= u^a_{ib'}\textsf{h}^{b'b}$ (see Paragraph \ref{annexeLie}).
Then, for $0\leq a,b\leq 3$ and $0\leq A\leq 9$, $\kappa^{ab}_A$ is defined by: $\kappa^{ab}_c=0$
for $0\leq c\leq 3$ and $\kappa^{ab}_i = 2u^{ab}_i$ for $4\leq i\leq 9$.
\item $\mathfrak{T}$ is the Abelian Lie group of translations on the Minkowski space, and
$\mathfrak{t}$ is its trivial Lie algebra, with basis $(t_0,t_1,t_2,t_3)$.
\item $\mathfrak{P} = \mathfrak{G}\ltimes \mathfrak{T}$ is the group of affine
isometries of $\mathbb{M}$ (or its universal cover), with Lie algebra
$\mathfrak{p} = \mathfrak{g}\oplus \mathfrak{t}$. We denote by
$(\mathfrak{l}_A)_{0\leq A\leq 9} = (t_0,\cdots,t_3,u_4\cdots,u_9)$, a basis of $\mathfrak{p}$.
If $\mathbb{M}$ is the Minkowski space, $\mathfrak{P}$ is the Poincar{\'e} Lie group.
\item $\mathfrak{g}^*$, $\mathfrak{t}^*$ and $\mathfrak{p}^*$ are the dual vector spaces
of respectively $\mathfrak{g}$, $\mathfrak{t}$ and $\mathfrak{p}$.
\item If $(e^0,e^1,e^2,e^3)$ is a coframe on a 4-dimensional manifold $\mathcal{X}$
(i.e. a collection of four 1-forms $e^0,e^1,e^2,e^3$ defined on an open subset of
$\mathcal{X}$ which is everywhere of rank 4) and if we denote by
$(\frac{\partial}{\partial e^0},\frac{\partial}{\partial e^1},\frac{\partial}{\partial e^2},\frac{\partial}{\partial e^3})$
the dual frame, we set
$e^{(4)}:= e^0\wedge e^1\wedge e^2\wedge e^3$ and
\[
e_a^{(3)}:= \frac{\partial}{\partial e^a}\iN e^{(4)},\quad
e_{ab}^{(2)}:=\frac{\partial}{\partial e^b}\iN e^{(4)}_a,\quad
e^{(1)}_{abc}:= \frac{\partial}{\partial e^c}\iN e^{(3)}_{ab}
\]
(note that $e^{(1)}_{abc} = \epsilon_{abcd}e^d$).
\item if $(e^0,\cdots,e^3,\gamma^4,\cdots,\gamma^9)$ is a coframe on a
10-dimensional manifold $\mathcal{P}$ and if 
$(\frac{\partial}{\partial e^0},\cdots,\frac{\partial}{\partial e^3},
\frac{\partial}{\partial \gamma^4},\cdots,\frac{\partial}{\partial \gamma^9})$ is its dual frame,
we set:
\[
\begin{array}{cccccc}
 e^{(4)} & := & e^0\wedge \cdots \wedge e^3, 
& \gamma^{(6)} & := & \gamma^4\wedge \cdots \wedge \gamma^9\\
 e^{(3)}_a & := & \frac{\partial}{\partial e^a}\iN e^{(4)},
& \gamma^{(5)}_i & := & \frac{\partial}{\partial \gamma^i}\iN \gamma^{(6)}\\
 e^{(2)}_{ab} & := & \frac{\partial}{\partial e^b}\iN e^{(3)}_a,
& \gamma^{(4)}_{ij} & := & \frac{\partial}{\partial \gamma^j}\iN \gamma^{(5)}_i\\
e^{(1)}_{abc} & := & \frac{\partial}{\partial e^c}\iN e^{(2)}_{ab},
& \gamma^{(3)}_{ijk} & := & \frac{\partial}{\partial \gamma^k}\iN \gamma^{(4)}_{ij}
\end{array}
\]
\item Similarly, if $(\alpha^0,\cdots,\alpha^3,\omega^4,\cdots,\omega^9)$ is another coframe on
$\mathcal{P}$, if $\alpha^{(4)}:= \alpha^0\wedge \alpha^1\wedge \alpha^2\wedge \alpha^3$,
$\omega^{(6)}:= \omega^4\wedge \cdots \wedge \omega^9$ and if
$(\frac{\partial}{\partial \alpha^0},\cdots,\frac{\partial}{\partial \alpha^3},
\frac{\partial}{\partial \omega^4},\cdots,\frac{\partial}{\partial \omega^9})$ is its dual frame
we use the same conventions:
$\alpha_a^{(3)}:= \frac{\partial}{\partial \alpha^a}\iN\alpha^{(4)}$,
$\alpha_{ab}^{(2)}:= \frac{\partial}{\partial \alpha^a}\wedge \frac{\partial}{\partial \alpha^b}\iN\alpha^{(4)}
= \frac{\partial}{\partial \alpha^b}\iN\alpha_a^{(3)}$, etc.,
$\omega_i^{(5)}:= \frac{\partial}{\partial \omega^i}\iN  \omega^{(6)}$,
 $\omega_{ij}^{(4)}:= \frac{\partial}{\partial \omega^i}\wedge \frac{\partial}{\partial \omega^j}\iN \omega^{(6)}
 = \frac{\partial}{\partial \omega^j}\iN \omega_i^{(5)}$, etc.
\end{itemize}

\section{The starting point of the approach}
Our first task consists in recasting the usual Weyl--Einstein--Cartan formulation of gravity
on the total space of the principal bundle of lorentzian frames on space-time in an invariant way.

\subsection{The Weyl--Einstein--Cartan action}
Consider a 4-dimensional manifold $\mathcal{X}$, the space-time. Dynamical fields
in the Weyl--Einstein--Cartan formulation
can be defined locally as being pairs $(e,A)$, where $e = (e^0,e^1,e^2,e^3)$ is a moving coframe on $\mathcal{X}$
(defining the metric $\textsf{h}_{ab}e^a\otimes e^b$ on the tangent bundle $T\mathcal{X}$)
and $A$ is a $\mathfrak{g}$-valued connection 1-form on $\mathcal{X}$.
The WEC (Weyl--Einstein--Cartan) action then reads 
\[
 \mathcal{A}_{EWC}[e,A] = \int_\mathcal{X}\frac{1}{2}\epsilon{_{abc}}^d e^a\wedge e^b\wedge (dA+A\wedge A){^c}_d
= \int_\mathcal{X}\frac{1}{2}\epsilon_{abcd}e^a\wedge e^b\wedge F^{cd},
\]
where $F := dA+A\wedge A$ and $F^{cd}:= F{^c}_{d'}\textsf{h}^{dd'}$.
Alternatively, by Lemma \ref{lemma1},
\[
 \mathcal{A}_{EWC}[e,A] = \int_\mathcal{X}e_{ab}^{(2)}\wedge F^{ab}
 = \int_\mathcal{X}u^{ab}_ie_{ab}^{(2)}\wedge F^i.
\]
It is possible to understand pairs $(e,A)$ in a more global and geometric way by assuming that a rank 4
vector bundle $V\mathcal{X}$ has been chosen over $\mathcal{X}$, equipped with a pseudo-metric $h$.
Then $A$ represents a connection of $V\mathcal{X}$ which respects the pseudo-metric $h$ and
$e$ represents a \emph{solder form}, i.e. a rank 4 section of the vector bundle over $\mathcal{X}$
whose fiber over $\textsf{x}\in \mathcal{X}$
is the set of linear maps from $T_\textsf{x}\mathcal{X}$ to $V_\textsf{x}\mathcal{X}$.
By choosing a family
of four local sections of $V\mathcal{X}$ that forms an orthonormal basis of $V\mathcal{X}$, we may decompose
locally $e$ and $A$ in terms of real valued 1-forms $e^a$ and $A{^c}_d$ and recover the previous description.
Note that this description still has the drawback that it rests on the
\emph{a priori} choice of a vector bundle $V\mathcal{X}$
over $\mathcal{X}$. This drawback will be removed in the model proposed in the following.

\subsection{Lifting to the principal bundle}\label{paragraphlifting}

It is well-known that the previous action is invariant by gauge transformations
of the form
\[
 (e,A)\longmapsto (g^{-1}e,g^{-1}dg + g^{-1}Ag),
\]
or, in indices,
\[
e^a\longmapsto (g^{-1}){^a}_{a'}e^{a'},
\quad A{^a}_b\longmapsto (g^{-1}){^a}_{a'}dg{^{a'}}_b + (g^{-1}){^a}_{a'}A{^{a'}}_{b'}g{^{b'}}_b
\]
where $g:\mathcal{X}\longrightarrow \mathfrak{G}$.
One way to picture geometrically this ambiguity is to lift
the variational problem on the \emph{total space} $\mathcal{P}$ of the principal bundle
of orthonormal frames on $V\mathcal{X}$ (with the right action of $\mathfrak{G}$ denoted by
$\mathcal{P}\times \mathfrak{G}\ni(\textsf{z},g)\longmapsto \textsf{z}\cdot g\in \mathcal{P}$).
This amounts roughly speaking to consider all possible gauge transformations
of a given field $(e,A)$ simultaneously. We then represent each pair $(e,A)$
by a pair of 1-forms $(\alpha,\omega)$
on $\mathcal{P}$ with values in the Poincar{\'e} Lie algebra $\mathfrak{p}$, i.e.
$\alpha$ takes values $\mathfrak{t}$ and $\omega$ takes values in $\mathfrak{g}$.

The price to pay however is that we need to assume that the $\mathfrak{p}$-valued
1-form $(\alpha,\omega)$
satisfies normalization and equivariance constraints. To write them, use the basis $(u_4,\cdots,u_9)$ of
$\mathfrak{g}$ and, for any $i = 4,\cdots, 9$, let $\rho_i$ be the tangent vector field on $\mathcal{P}$
induced by the right action of $u_i$ on $\mathcal{P}$. Indeed we assume that the lift $\omega$
of $A$ satisfies the following \emph{normalization} and \emph{equivariance} properties
respectively (see \cite{helein14})
\begin{equation}\label{normalisationStandard}
 \rho_i\iN  \omega = u_i,
\end{equation}
\begin{equation}\label{equivarianceStandard}
 L_{\rho_i}\omega + [u_i,\omega] = 0,
\end{equation}
where $L_{\rho}$ denotes the Lie derivative with respect to a vector field $\rho$. Similarly $\alpha$ satisfies
respectively the \emph{normalization} and \emph{equivariance} properties
\begin{equation}\label{normalisationSoudureStandard}
 \rho_i\iN\alpha = 0,
\end{equation}
\begin{equation}\label{equivarianceSoudureStandard}
 L_{\rho_i}\alpha + u_i\alpha = 0.
\end{equation}

The relationship with the previous description is as follows: for any $\mathfrak{p}$-valued 1-form
$(\alpha,\omega)$ on $\mathcal{P}$
which satisfies (\ref{normalisationStandard}), (\ref{equivarianceStandard}), (\ref{normalisationSoudureStandard})
and (\ref{equivarianceSoudureStandard}) and for any local section
$\sigma:\mathcal{X}\longrightarrow \mathcal{P}$, we obtain a pair $(e,A)$ on $\mathcal{X}$ simply by setting
$e=\sigma^*\alpha$ and $A=\sigma^*\omega$.

Conversely, given a pair $(e,A)$ on $\mathcal{X}$ and a local section
$\sigma:\mathcal{X}\longrightarrow \mathcal{P}$, this provides us with a local trivialization
\[
 \begin{array}{cccl}
  \mathcal{T}: & \mathcal{P} & \longrightarrow &  \mathcal{X}\times \mathfrak{G}\\
& \textsf{z} & \longmapsto & (\textsf{x},g)
 \end{array}
\]
where $(\textsf{x},g)$ is s.t. $\textsf{z} = \sigma(\textsf{x})\cdot g$. We can then
associate to $(e,A)$ a $\mathfrak{p}$-valued 1-form $(\alpha,\omega)$ on $\mathcal{P}$
which satisfies (\ref{normalisationStandard}),
(\ref{equivarianceStandard}), (\ref{normalisationSoudureStandard})
and (\ref{equivarianceSoudureStandard}) given by $\alpha = \mathcal{T}^*(g^{-1}e)$
and $\omega = \mathcal{T}^*(g^{-1}Ag+g^{-1}dg)$ (see \cite{helein14}).

Lastly let us define $\gamma:= \mathcal{T}^*(g^{-1}dg)$ and denote by
$\gamma^4,\cdots,\gamma^9$ the components of $\gamma$
in the basis $(u_4,\cdots, u_9)$, i.e. s.t. $\gamma = u_i\gamma^i$. We can lift the
action $\mathcal{A}_{EWC}$
to a functional on the space of $\mathfrak{p}$-valued 1-forms $(\alpha,\omega)$ by setting:
\[
 \widehat{\mathcal{A}}_{EWC}[\alpha,\omega] =
\int_\mathcal{P}\frac{1}{2}\epsilon{_{abc}}^d \alpha^a\wedge \alpha^b\wedge
(d\omega+\omega\wedge \omega){^c}_d \wedge \gamma^1\wedge \cdots \wedge \gamma^6.
\]
Alternatively, by setting $\alpha^{(2)}_{ab}:= \frac{1}{2}\epsilon_{abcd}\alpha^c\wedge \alpha^d$,
$\Omega:= d\omega+\omega\wedge \omega$,
$\Omega^{ab}:= \Omega{^a}_{b'}\textsf{h}^{bb'}$
and $\gamma^{(6)}:= \gamma^4\wedge \cdots \wedge \gamma^9$, we can write
\begin{equation}\label{6bis}
 \widehat{\mathcal{A}}_{EWC}[\alpha,\omega] = \int_\mathcal{P}
\alpha^{(2)}_{ab}\wedge \Omega^{ab}\wedge \gamma^{(6)}
= \int_\mathcal{P}u^{ab}_i\alpha^{(2)}_{ab}\wedge \Omega^i\wedge \gamma^{(6)}.
\end{equation}
Then critical points of $\mathcal{A}_{EWC}$ correspond to critical points of $\widehat{\mathcal{A}}_{EWC}$
under the constraints (\ref{normalisationStandard}),
(\ref{equivarianceStandard}), (\ref{normalisationSoudureStandard})
and (\ref{equivarianceSoudureStandard}).

\subsection{Forgetting the fibration}\label{paragraphforget}
A key step for our purpose is to translate the previous conditions on $(\alpha,\omega)$
in a situation where the fibration $\mathcal{P}\longrightarrow \mathcal{X}$ is not given a priori.
For that we claim that the normalization conditions (\ref{normalisationStandard})
and (\ref{normalisationSoudureStandard}) are not essential (this will be confirmed by the following).
We hence translate the equivariance conditions (\ref{equivarianceStandard}) and
(\ref{equivarianceSoudureStandard}) without reference to the normalization conditions.

We first observe that, if (\ref{normalisationStandard}) holds, then
$L_{\rho_i}\omega = \rho_i\iN d\omega + d(\rho_i\iN \omega)
= \rho_i\iN d\omega + du_i = \rho_i\iN d\omega$ and
$\rho_i\iN \omega\wedge \omega = [\omega(\rho_i),\omega] = [u_i,\omega]$; hence the
l.h.s. of (\ref{equivarianceStandard}) is equal to $L_{\rho_i}\omega + [u_i,\omega]
= \rho_i\iN d\omega + \rho_i\iN \omega\wedge \omega$. Thus, assuming (\ref{normalisationStandard}),
(\ref{equivarianceStandard}) is equivalent to
\begin{equation}\label{equivarianceNouvelle0}
\rho_i\iN (d\omega + \omega\wedge \omega) = 0,\quad \forall i = 1,\cdots, 6.
\end{equation}
Similarly, if (\ref{normalisationStandard}) and (\ref{normalisationSoudureStandard}) hold,
$L_{\rho_i}\alpha = \rho_i\iN d\alpha + d(\rho_i\iN \alpha)
= \rho_i\iN d\alpha + d0 = \rho_i\iN d\alpha$ and $\rho_i\iN \omega\wedge \alpha = [u_i,\alpha]$.
Hence, if we assume (\ref{normalisationStandard}) and (\ref{normalisationSoudureStandard}),
(\ref{equivarianceSoudureStandard}) is equivalent to
\begin{equation}\label{equivarianceSoudureNouvelle0}
\rho_i\iN (d\alpha + \omega\wedge \alpha) = 0,\quad \forall i = 1,\cdots, 6.
\end{equation}
Now both equations (\ref{equivarianceNouvelle0}) and (\ref{equivarianceSoudureNouvelle0}) are linear
in $\rho_i$ and so are also valid if we replace $\rho_i$ by any tangent vector field
$\rho$ on $\mathcal{P}$ which is a linear
combination of $\rho_4,\cdots,\rho_9$. Such vector fields are tangent to the fibers of $\mathcal{P}
\longrightarrow \mathcal{X}$ or, equivalentely, are characterized by the property
$\rho\iN \alpha^a = 0$, $\forall a = 0,\cdots,3$.
Hence (\ref{equivarianceNouvelle0}) and (\ref{equivarianceSoudureNouvelle0})
are equivalent to the implication $[\rho\iN \alpha=0]\Longrightarrow
[\rho\iN (d\omega + \omega\wedge \omega) = \rho\iN (d\alpha + \omega\wedge \alpha) = 0]$.
This is also equivalent to claim that there exists functions $Q{^{a}_{}}_{bcd}$ and $Q{^{a}_{}}_{cd}$
on $\mathcal{P}$ s.t.
\begin{equation}\label{equivarianceNouvelle1}
 (d\alpha +\omega\wedge\alpha)^{a} =
\frac{1}{2}Q{^{a}_{}}_{cd}\alpha^c\wedge \alpha^d
\quad \hbox{and}\quad
(d\omega +\omega\wedge\omega){^{a}_{}}_{b} = 
\frac{1}{2}Q{^{a}_{}}_{bcd}\alpha^c\wedge \alpha^d.
\end{equation}
Note that, if we set $Q{^{ab}_{}}_{cd}:= Q{^{a}_{}}_{b'cd}\textsf{h}^{b'b}$, we have
$Q{^{ab}_{}}_{cd} + Q{^{ba}_{}}_{cd} =0$ and that
we may assume w.l.g. that $Q{^{a}_{}}_{bcd} + Q{^{a}_{}}_{bdc} = Q{^{a}_{}}_{cd} + Q{^{a}_{}}_{dc} = 0$.

Now let us return to the action.
A key observation is that, since $\omega = \gamma + \mathcal{T}^*(g^{-1}Ag)
$ and since $\mathcal{T}^*(g^{-1}Ag)$ is a linear combination of
$\alpha^0,\alpha^1,\alpha^2$ and $\alpha^3$, we have
\begin{equation}\label{9bis}
 \alpha^{(4)}\wedge \gamma^4\wedge \cdots \wedge \gamma^9
= \alpha^{(4)}\wedge \omega^4\wedge \cdots \wedge \omega^9,
\quad \forall c,d,
\end{equation}
where the $\omega^i$ are the coefficients of the decomposition of $\omega$
in the basis $(u_4,\cdots, u_9)$.

But if we assume that (\ref{equivarianceNouvelle1}) is satisfied we have
$\Omega^{ab} =  \frac{1}{2}Q{^{ab}_{}}_{cd}\alpha^c\wedge \alpha^d$ and
hence $\alpha^{(2)}_{ab}\wedge \Omega^{ab} = Q{^{ab}_{}}_{ab}\alpha^{(4)}$
(see Lemma \ref{lemma0}). Hence, by using (\ref{6bis}) and (\ref{9bis}), it follows
that, \emph{if $(\alpha,\omega)$ satisfies (\ref{equivarianceNouvelle1})},
\begin{equation}\label{action-libre-0}
 \widehat{\mathcal{A}}_{EWC}[\alpha,\omega] = \int_\mathcal{P}
\alpha^{(2)}_{ab}\wedge \Omega^{ab}\wedge \omega^4\wedge \cdots \wedge \omega^9
= \int_\mathcal{P}u^{ab}_i\alpha^{(2)}_{ab}\wedge \Omega^i\wedge \omega^{(6)},
\end{equation}
where $\omega^{(6)}:= \omega^4\wedge \cdots \wedge \omega^9$.
Thus we are led to study critical points of the action defined in (\ref{action-libre-0}) under
the constraints (\ref{equivarianceNouvelle1}). As in \cite{helein14} such constraints are non-holonomic
and thus a source of difficulties. We will follow a similar approach to the one in \cite{helein14} and perform
a Legendre transform of the former variational problem within the multisymplectic framework.

\section{Towards a multisymplectic formulation}\label{sectionmultipatates}

\subsection{The canonical 1-form on $\mathfrak{p}\otimes T^*\mathcal{P}$}
In order to facilitate the computation, we introduce the vector bundle $\mathfrak{p}\otimes T^*\mathcal{P}$
over $\mathcal{P}$, whose fiber at point $\textsf{z}\in \mathcal{P}$ is the tensor product
$\mathfrak{p}\otimes T^*_\textsf{z}\mathcal{P}$ and can be canonically identified with the
space of linear maps from $T_\textsf{z}\mathcal{P}$ to the Poincar{\'e} Lie algebra $\mathfrak{p}$.
A point in $\mathfrak{p}\otimes T^*\mathcal{P}$ will be denoted by $(\textsf{z},\textsf{y})$,
where $\textsf{z}\in \mathcal{P}$ and $\textsf{y}\in\mathfrak{p}\otimes T^*_\textsf{z}\mathcal{P}$.
This bundle is equipped with the canonical $\mathfrak{p}$-valued 1-form $\eta$
(a section of $\mathfrak{p}\otimes T^*(\mathfrak{p}\otimes T^*\mathcal{P})$) defined by
\[
 \forall (\textsf{z},\textsf{y})\in \mathfrak{p}\otimes T^*\mathcal{P},
\forall v\in T_{(\textsf{z},\textsf{y})}(\mathfrak{p}\otimes T^*\mathcal{P}),\quad
\eta_{(\textsf{z},\textsf{y})}(v) = \textsf{y}(d\pi_{(\textsf{z},\textsf{y})}(v)),
\]
where $\pi = \pi_{\mathfrak{p}\otimes T^*\mathcal{P}}:
\mathfrak{p}\otimes T^*\mathcal{P}\longrightarrow \mathcal{P}$ is the canonical projection
map. This $\mathfrak{p}$-valued 1-form can be decomposed as
$\eta = \mathfrak{l}_A\eta^A$, where each $\eta^A$ is a 1-form on $\mathcal{P}$.

We introduce the following coordinates on $\mathfrak{p}\otimes T^*\mathcal{P}$:
\begin{itemize}
 \item  $(z^I)_{1\leq I\leq 10}$ are local coordinates on $\mathcal{P}$; thus they provide us with
 locally defined functions $z^I\circ \pi$ on $\mathfrak{p}\otimes T^*\mathcal{P}$. In the following we
 write abusively $z^I\simeq z^I\circ \pi$.
 \item for any $\z\in \mathcal{P}$, we can define the coordinates
$(\eta^A_I)_{0\leq A\leq 9;1\leq I\leq 10}$ on the space  $\mathfrak{p}\otimes T^*_\z\mathcal{P}$
in the basis $(\mathfrak{l}_A\otimes dz^I)_{0\leq A\leq 9;1\leq I\leq 10}$.
\end{itemize}
Hence $\mathfrak{p}\otimes T^*\mathcal{P}$ is endowed with local coordinates $(z^I,\eta^A_I)$. In these coordinates
$\eta$ reads
\[
 \eta = \mathfrak{l}_A\eta^A_Idz^I.
\]
We may split $\eta = \etazero + \etaun$, according to the decomposition
$\mathfrak{p} = \mathfrak{g} \oplus \mathfrak{t}$. Note that
$\etaun = \eta^a\mathfrak{l}_a = \eta^at_a$, where $0\leq a\leq 3$,
and $\etazero = \eta^i\mathfrak{l}_i = \eta^iu_i$, where $4\leq i\leq 9$.
We also set  $\etazero{^a}_b = u^a_{ib}\etazero{^i}$.
Any pair $(\alpha,\omega)$ as considered in the previous section is a section of
$\mathfrak{p}\otimes T^*\mathcal{P}$ over $\mathcal{P}$. In the following we identify such a pair with
a map $\varphi$ from $\mathcal{P}$ to the total space of $\mathfrak{p}\otimes T^*\mathcal{P}$ 
(a manifold of dimension 110) such that
$\pi\circ \varphi(\textsf{z}) = \textsf{z}$, $\forall \textsf{z}\in \mathcal{P}$, by letting
\begin{equation}\label{identification-forme-section}
 (\alpha,\omega) = \varphi^*\eta
\quad \hbox{or}\quad
\alpha = \varphi^*\etaun \hbox{ and } \omega = \varphi^*\etazero.
\end{equation}
As for $\etaun$ and $\etazero$ we denote by $(\omega^i)_{1\leq i\leq 6}$ the components of the decomposition 
$\omega = u_i\omega^i$ and we set $\omega{^a}_b = u^a_{ib}\omega^i$; similarly we write
$(\alpha^a)_{0\leq a\leq 3}$ the components of $\alpha$.

We now recast the action $\widehat{\mathcal{A}}_{EWC}$ as follows.
We define the following 10-form on $\mathfrak{p}\otimes T^*\mathcal{P}$
(i.e. a section of $\Lambda^{10}T^*(\mathfrak{p}\otimes T^*\mathcal{P})$):
\begin{equation}\label{lagrangienPalatiniStart}
 \mathcal{L} = u^{ab}_i \etaun{_{ab}}^{(2)}\wedge (d\etazero+\etazero\wedge\etazero)^i
 \wedge \etazero{^{(6)}},
\end{equation}
where $\etaun{_{ab}}^{(2)}:= \frac{1}{2}\epsilon_{abcd}\eta^c\wedge \eta^d$
and $\etazero{^{(6)}}:= \eta^4\wedge \cdots \wedge \eta^9$.
Note that the definition of $\mathcal{L}$
does not require a fibration on $\mathcal{P}$ over some manifold $\mathcal{X}$: it is canonically defined 
on any manifold of the form $\mathfrak{p}\otimes T^*\mathcal{P}$, where $\mathcal{P}$ is any
10-dimensional manifold. We can now give another expression for the action
(\ref{action-libre-0}):
\begin{equation}\label{action-libre-1}
 \widehat{\mathcal{A}}_{EWC}[\alpha,\omega] = \int_\mathcal{P}
\varphi^*\mathcal{L},
\end{equation}
where $\varphi$ is such that (\ref{identification-forme-section}) holds.

The constraints (\ref{equivarianceNouvelle1}) then translate as the following conditions on
$\varphi$:
\begin{equation}\label{equivarianceNouvelle2e}
 \exists Q{^{a}_{}}_{cd}\in \mathcal{C}^\infty(\mathcal{P}),\quad
 \quad
(d\alpha +\omega\wedge\alpha)^{a} =
\frac{1}{2}Q{^{a}_{}}_{cd}\alpha^c\wedge \alpha^d.
\end{equation}
\begin{equation}\label{equivarianceNouvelle2A}
 \exists Q{^{a}_{}}_{bcd}\in \mathcal{C}^\infty(\mathcal{P}),\quad
 (d\omega +\omega\wedge\omega){^{a}_{}}_{b} = 
\frac{1}{2}Q{^{a}_{}}_{bcd}\alpha^c\wedge \alpha^d,
\end{equation}
Conditions (\ref{equivarianceNouvelle2e}) and
(\ref{equivarianceNouvelle2A}) are equivalent to 
\begin{equation}\label{equivarianceNouvelle2}
 \exists Q{^A_{}}_{cd}\in \mathcal{C}^\infty(\mathcal{P}),\quad
 \varphi^*(d\eta +\frac{1}{2}[\eta\wedge\eta])^A = 
\frac{1}{2}Q{^A_{}}_{cd}\varphi^*(\eta^c\wedge \eta^d)
\end{equation}
(compare with (\ref{equivarianceNouvelle3}) below).

\subsection{The Poincar{\'e}--Cartan form $\theta_{Tot}$}
Among the many possible multisymplectic manifolds, we need to choose a convenient one as
a framework for the Legendre transform of our problem, i.e. a suitable submanifold of 
the manifold\footnote{The $\left(110 + \frac{110!}{100!10!}\right)$-dimensional
universal Lepage--Dedecker manifold $\Lambda^{10}T^*(\mathfrak{p}\otimes T^*\mathcal{P})$
is far too big.} $\Lambda^{10}T^*(\mathfrak{p}\otimes T^*\mathcal{P})$.
Inspired by \cite{helein14} we choose the total space of the fiber bundle over $\mathcal{P}$
\[
 \mathcal{M}_{Tot}:= \R\oplus_\mathcal{P}\left(\mathfrak{p}^*\otimes\Lambda^8T^*\mathcal{P}\right)
\oplus_\mathcal{P}\left(\mathfrak{p}\otimes T^*\mathcal{P}\right).
\]
We introduce the following coordinates on $\mathcal{M}_{Tot}$:
\begin{itemize}
 \item we extend in a natural way the coordinates $(z^I,\eta^A_I)$ on $\mathfrak{p}\otimes T^*\mathcal{P}$
to functions on $\mathcal{M}_{Tot}$.
 \item we let $(\mathfrak{l}^A)_{0\leq A\leq 9}$ be the basis of $\mathfrak{p}^*$ which is dual
 to $(\mathfrak{l}_A)_{0\leq A\leq 9}$; 
 for any $\z\in \mathcal{P}$, let $dz^{(10)}:= dz^1\wedge\cdots \wedge dz^{10}$ and
 $dz^{(8)}_{IJ}:= \frac{\partial}{\partial z^J}\iN \frac{\partial}{\partial z^I}\iN dz^{(10)}$.
 We define the coordinates $\psi_A^{IJ} = - \psi_A^{JI}$ on the space
 $\mathfrak{p}^*\otimes \Lambda^8T^*_\z\mathcal{P}$ in the basis
 $(\mathfrak{l}^A\otimes dz^{(8)}_{IJ})_{0\leq A\leq 9;1\leq I<J\leq 10}$.
 Then $\mathfrak{p}^*\otimes \Lambda^8T^*\mathcal{P}$ is endowed with local coordinates $(z^I,\psi_A^{IJ})$.
 \item endow the real line $\R$ with the coordinate $h$.
\end{itemize}
Then a complete system of coordinates on $\mathcal{M}_{Tot}$ is $(z^I, h, \eta^A_I,\psi_A^{IJ})$.

On $\mathfrak{p}^*\otimes \Lambda^8T^*\mathcal{P}$ is also defined a canonical $\mathfrak{p}^*$-valued
8-form $\psi$ defined by: $\forall (\textsf{z},\textsc{m})\in \mathfrak{p}^*\otimes \Lambda^8T^*\mathcal{P}$,
\[
\forall w_1,\cdots,w_8\in T_{(\textsf{z},\textsc{m})}(\mathfrak{p}^*\otimes \Lambda^8T^*\mathcal{P}),\quad
\psi_{(\textsf{z},\textsc{m})}(w_1,\cdots,w_8) = \textsc{m}(d\pi_{(\textsf{z},\textsc{m})}(w_1),\cdots,
d\pi_{(\textsf{z},\textsc{m})}(w_8)),
\]
where $\pi = \pi_{\mathfrak{p}^*\otimes \Lambda^8T^*\mathcal{P}}: 
\mathfrak{p}^*\otimes \Lambda^8T^*\mathcal{P}\longrightarrow \mathcal{P}$ is
the canonical projection map. This $\mathfrak{p}^*$-valued 8-form decomposes as
$\psi = \psi_A\mathfrak{l}^A$.
In local coordinates $(z^I,\psi_A^{IJ})$ $\psi$ reads
\[
 \psi = \frac{1}{2}\mathfrak{l}^A\psi_A^{IJ}dz^{(8)}_{IJ}.
\]
We now define define the Poincar{\'e}--Cartan 10-form on $\mathcal{M}_{Tot}$
\[
 \theta_{Tot}:= h\eta^{(10)} + \psi_A\wedge (d\eta + \frac{1}{2}[\eta\wedge\eta])^A,
\]
where $\eta^{(10)}:= \eta^1\wedge \cdots\wedge \eta^{10}$
Alternatively, 
\[
 \theta_{Tot}:= h\eta^{(10)} + \psi_a\wedge (d\etaun +\etazero\wedge\etaun)^a
+ \psi_i\wedge (d\etazero +\etazero\wedge\etazero)^i.
\]

\subsection{The first jet bundle on $\mathfrak{p}\otimes T^*\mathcal{P}$}
We now need to introduce the first jet bundle of the bundle $\mathfrak{p}\otimes T^*\mathcal{P}$
over $\mathcal{P}$, which plays a role analogue to the tangent bundle in Mechanics.
Recall that a section $\varphi$ of the fiber bundle $\mathfrak{p}\otimes T^*\mathcal{P}$ 
can be seen as a map
$\varphi: \mathcal{P}\longrightarrow \mathfrak{p}\otimes T^*\mathcal{P}$ such that
$\pi_{\mathfrak{p}\otimes T^*\mathcal{P}}\circ \varphi = \hbox{Id}_\mathcal{P}$. Such a section is completely
characterized by the functions $\eta^A_I\circ \varphi$.
The jet space $J^1(\mathcal{P},\mathfrak{p}\otimes T^*\mathcal{P})$
is the manifold of triplets $(\z,\y,\ypoint)$, where
$(\z,\y)\in \mathfrak{p}\otimes T^*\mathcal{P}$ and
$\ypoint$ is the equivalence class of local sections $\varphi$ of $\mathfrak{p}\otimes T^*\mathcal{P}$
over a neighborhood of $\z$ such that $\varphi(\z) = \y$, for the equivalence relation:
$\varphi_1\simeq \varphi_2$ iff $d(\eta^A_I\circ \varphi_1)_\z = d(\eta^A_I\circ \varphi_2)_\z$,
$\forall I,A$. We then write $[\varphi]_{\z,\y}$ the class of $\varphi$.
Local coordinates on  $J^1(\mathcal{P},\mathfrak{p}\otimes T^*\mathcal{P})$
are $(z^I,\eta^A_I,\eta^A_{I;J})$, where
\[
 \eta^A_{I;J}(\ypoint) = \frac{\partial (\eta^A_I\circ \varphi)}{\partial z^J}(\z)
 \quad\hbox{where}\quad 
 \ypoint = [\varphi]_{\z,\y},
\]
or alternatively
\[
 \eta^A_{I;J}(\ypoint)dz^J = d(\eta^A_I\circ \varphi)_\z = (\varphi^*d\eta^A_I)_\z.
\]
It will be however convenient to introduce the families of functions
$(S^A_{IJ})_{0\leq A\leq 9;1\leq I,J\leq 10}$ and $(A{^A}_{BC})_{0\leq A,B,C\leq 9}$
on $J^1(\mathcal{P},\mathfrak{p}\otimes T^*\mathcal{P})$, defined respectively by
\[
 S^A_{IJ}(\ypoint) = \frac{1}{2}\left( \eta^A_{J;I}(\ypoint) +\eta^A_{I;J}(\ypoint)\right)
\]
(note that $S^A_{JI} = S^A_{IJ}$)
and, for $A{^A}_{BC}$, by the conditions $A{^A}_{BC}+A{^A}_{CB}=0$ and:
\[
 \frac{1}{2}A{^A}_{BC}(\ypoint)\varphi^*(\eta^B\wedge \eta^C)_\z = 
 \varphi^*\left(d\eta^A + \frac{1}{2}\left[\eta\wedge \eta\right]^A\right)_\z.
\]
We remark that
\[
 \eta^A_{I;J}(\ypoint) = S^A_{IJ}(\ypoint) - \frac{1}{4}A{^A}_{BC}(\ypoint)
 \left|\begin{array}{cc}
                           \eta^B_I(\y) & \eta^B_J(\y)\\
                           \eta^C_I(\y) & \eta^C_J(\y)
                          \end{array}\right|
                          + \frac{1}{2}[\eta_I(\y);\eta_J(\y)]^A.
\]
Hence a system of coordinates on $\mathfrak{p}\otimes T^*\mathcal{P}$ is:
\[
 (z^I)_{1\leq I\leq 10}, \quad (\eta^A_I)_{0\leq A\leq 9;1\leq I\leq 10},
\quad (S^A_{IJ})_{0\leq A\leq 9;1\leq I\leq J\leq 10}
\quad \hbox{and}\quad  (A{^A}_{BC})_{0\leq A\leq 9;0\leq B<C\leq 9}.
\]
In fact, all relevant quantities (the constraints, the Lagrangian density and the Poincar{\'e}--Cartan
form) depend only on  $z^I,\eta^A_I$ and $A{^A}_{BC}$ (and not on the $S^A_{IJ}$'s).
Indeed for instance the pull-back of $d\eta+\frac{1}{2}[\eta\wedge \eta]$ by any section $\varphi$ has the a priori
decomposition
\[
 \varphi^*(d\eta+\frac{1}{2}[\eta\wedge \eta])^A = \frac{1}{2}A{^A_{}}_{cd}\alpha^c\wedge \alpha^d
+ A{^A_{}}_{ck}\alpha^c\wedge \omega^k + \frac{1}{2}A{^A_{}}_{jk}\omega^j\wedge \omega^k,
\]
so that (\ref{equivarianceNouvelle2}) amounts to impose that
\begin{equation}\label{equivarianceNouvelle3}
\exists Q{^A}_{cd}\in \mathcal{C}^\infty(\mathcal{P}),\quad
A{^A_{}}_{cd} = Q{^A_{}}_{cd}(\z)\quad \hbox{and}\quad 
 A{^A_{}}_{ck} = A{^A_{}}_{jk} = 0,\quad \forall A,c,d,j,k.
\end{equation}

\subsection{The Legendre transform}
Let $(\z,\y,\ypoint)\in J^1(\mathcal{P},\mathfrak{p}\otimes T^*\mathcal{P})$
and let $\varphi$ be a section such that $[\varphi]_{\z,\y} = \ypoint$. In order to compute
the Legendre transform at $(\z,\y,\ypoint,h,p)$ we need to evaluate $\varphi^*(\theta_{Tot} - \mathcal{L})$ and
to determine the value of the quantity $W(\z,\y,\ypoint,h,p)$ which is defined by
$\varphi^*(\theta_{Tot} - \mathcal{L}) = W(\z,\y,\ypoint,h,p)\varphi^*(\alpha^{(4)}\wedge \omega^{(6)})$
(see \cite{hk2} for details).

\subsubsection{Computation of $\varphi^*\theta_{Tot}$}
We decompose\footnote{Beware that sign conventions below are different from \cite{helein14}.}
\begin{eqnarray}
 \psi_a & = & \frac{1}{2}\psi{_a}{^{cd}}\alpha_{cd}^{(2)}\wedge \omega^{(6)}
 - \psi{_a}{^{ck}}\alpha_c^{(3)}\wedge \omega_k^{(5)}
 + \frac{1}{2}\psi{_a}{^{jk}}\alpha^{(4)}\wedge \omega_{jk}^{(4)}\label{decompostionpa}\\
\psi_i & = &  \frac{1}{2}\psi{_i}{^{cd}}\alpha_{cd}^{(2)}\wedge \omega^{(6)}
 - \psi{_i}{^{ck}}\alpha_c^{(3)}\wedge \omega_k^{(5)}
 + \frac{1}{2}\psi{_i}{^{jk}}\alpha^{(4)}\wedge \omega_{jk}^{(4)}\label{decompostionpi}
\end{eqnarray}
Moreover the pull-back of $\theta_{Tot}$ by a section
$\varphi:\mathcal{P}\longrightarrow \mathfrak{p}\otimes T^*\mathcal{P}$
reads
\[
  \varphi^*\theta_{Tot}  = (h\circ \varphi)\varphi^*\eta^{(10)}
   + (\varphi^*\psi_a)\wedge (d\alpha+\omega\wedge \alpha)^a
+ (\varphi^*\psi_i)\wedge (d\omega+\omega\wedge \omega)^i.
\]
Hence, in view of the constraints (\ref{equivarianceNouvelle2e}) and (\ref{equivarianceNouvelle2A})
and of Lemma \ref{lemma0}, this  gives us
\[
 \varphi^*\theta_{Tot}  = \left[(h\circ \varphi) + \frac{1}{2}(\psi{_a}{^{cd}}\circ \varphi)Q{^a}_{cd}
  + \frac{1}{2}(\psi{_i}{^{cd}}\circ \varphi)Q{^i}_{cd}\right]\varphi^*\eta^{(10)},
\]
for some functions $Q{^a}_{cd}$ and $Q{^i}_{cd}$ which depends on $\varphi$.

\subsubsection{Computation of $\varphi^*\mathcal{L}$}
Using Formula (\ref{lagrangienPalatiniStart}) for $\mathcal{L}$ and
the constraints (\ref{equivarianceNouvelle2e}) and (\ref{equivarianceNouvelle2A}) we find that
\[
 \varphi^*\mathcal{L} = \left(u_i^{ab}\alpha^{(2)}_{ab}\wedge
 \frac{1}{2}Q{^i}_{cd}\alpha^c\wedge \alpha^d\right) \wedge \omega^{(6)}
 = u_i^{ab}Q{^i}_{ab}\varphi^*\eta^{(10)}.
\]
Hence 
\[
 \varphi^*\left(\theta_{Tot}-\mathcal{L}\right) =
 \left[(h\circ \varphi) + \frac{1}{2}(\psi{_a}{^{cd}}\circ \varphi)Q{^a}_{cd}
  + \left(\frac{1}{2}\psi{_i}{^{cd}}\circ \varphi - u_i^{cd}\right)Q{^i}_{cd}
\right]\varphi^*\eta^{(10)}.
\]
Note that this form takes into account the constraints imposed on $\ypoint$.

\subsubsection{Conclusion: the Legendre transform}
From the following we deduce that
\begin{equation}\label{enfinW}
 W(\z,\y,\ypoint,h,p) = (h\circ \varphi)
+ \frac{1}{2}(\psi{_a}{^{cd}}\circ \varphi)A{^a}_{cd}
+ \left(\frac{1}{2}\psi{_i}{^{cd}}\circ \varphi - u_i^{cd}\right)A{^i}_{cd}.
\end{equation}
The Legendre correspondence holds on the points with coordinates
$(h,\z,\y,\ypoint,\psi)$ which are critical points of $W$ with
respect to infinitesimal variations of $\ypoint$ which respect the constraints, i.e., such that
\[
 \frac{\partial W}{\partial A{^A}_{bc}}  = 0\quad \hbox{and}\quad
\frac{\partial W}{\partial S^A_{IJ}} = 0.
\]
The second relation is trivially satisfied and the first one is equivalent to:
\begin{equation} \label{correspondanceLegendre}
 \psi{_a}{^{cd}}\circ \varphi = 0 \quad\hbox{and}\quad  \psi{_i}{^{cd}}\circ \varphi = 2u_i^{cd}.
\end{equation}
The value of the Hamiltonian function is then the restriction of $W$ at the points where
(\ref{correspondanceLegendre}) holds, i.e. simply:
\begin{equation}\label{valeurHamiltonien}
 H(\z,\y,h,p) = h.
\end{equation}
Our final multisymplectic manifold will be the submanifold $\mathcal{M}$ of $\mathcal{M}_{Tot}$
which is the intersection of the image of the Legendre correspondence ---precisely defined
by the constraints (\ref{correspondanceLegendre})--- with the hypersurface
$h=0$. By denoting $\theta$ the restriction of $\theta_{Tot}$ to $\mathcal{M}$:
\begin{equation}\label{nouveauTheta}
\begin{array}{ccr}
 \theta & = & \displaystyle \left(-\psi{_a}{^{ck}}{\etaun}_c^{(3)}\wedge {\etazero}_k^{(5)}
 + \frac{1}{2}\psi{_a}{^{jk}}{\etaun}^{(4)}\wedge {\etazero}_{jk}^{(4)}\right)
\wedge (d{\etaun}+{\etazero}\wedge {\etaun})^a
\\
 &  & \displaystyle + \left(u_i^{cd}{\etaun}_{cd}^{(2)}\wedge {\etazero}^{(6)}
 - \psi{_i}{^{ck}}{\etaun}_c^{(3)}\wedge {\etazero}_k^{(5)}
 + \frac{1}{2}\psi{_i}{^{jk}}{\etaun}^{(4)}\wedge {\etazero}_{jk}^{(4)}\right)
\wedge (d{\etazero}+{\etazero}\wedge{\etazero})^i
\end{array}
\end{equation}
Note that taking into account that $\eta$ and $\psi$ are respectively $\mathfrak{p}$- and $\mathfrak{p}^*$-valued,
our Poincar{\'e}--Cartan form has the simple structure:
\begin{equation}\label{thetaDeBase}
\theta:=  \psi\wedge (d\eta+\frac{1}{2}[\eta\wedge \eta]),
\end{equation}
where the duality pairing between coefficients of $\psi$ and $\eta$ is implicitely assumed.


\section{The Hamilton equations}\label{sectionHamilton}
Let $\kappa_A^{cd}$ be defined  for $A=a$ and $A=i$ by:
\[
  \kappa_a^{cd} := 0
 \quad \hbox{and} \quad \kappa_i^{cd} := 2u_i^{cd}.
\]
We can summarize the previous computation as follows: we work in the manifold $\mathcal{M}$
which can be identified with the submanifold of 
$\left(\mathfrak{p}^*\otimes\Lambda^8T^*\mathcal{P}\right)
\oplus_\mathcal{P}\left(\mathfrak{p}\otimes T^*\mathcal{P}\right)$
defined by the equations
\begin{equation}\label{constraints4}
 \psi{_A}^{cd} = \kappa_A^{cd}.
\end{equation}
or equivalentely by
\begin{equation}\label{contrainteVariationnelle}
 {\etaun}{}^c\wedge{\etaun}{}^d\wedge \psi_A = \kappa_A^{cd}\eta^{(10)},\quad \forall A,c,d,
\end{equation}
The manifold $\mathcal{M}$ will be our multisymplectic phase space: it is endowed with the
pre-multisymplectic 11-form $d\theta$. Solutions of the Hamilton equations can be described
as being 10-dimensional oriented submanifolds $\Gamma$ of $\mathcal{M}$ which satisfy the
\emph{independence condition}
\begin{equation}\label{independence}
 \eta^{(10)}|_\Gamma \neq 0
\end{equation}
and the Hamilton--Volterra--De Donder--Weyl (HVDW) equations
\begin{equation}\label{hvdw0}
 \forall \textsc{m}\in \Gamma,\forall \xi\in T_\textsc{m}\mathcal{M},\quad
(\xi\iN d\theta)|_{T_\textsc{m}\Gamma} = 0.
\end{equation}

\subsection{The solutions as critical points of an action functional}
In order to determine Equation (\ref{hvdw0}) we will use the fact that it is also the
Euler--Lagrange equations satisfied by the
critical points of the functional $\mathcal{A}[\Gamma]:=\int_\Gamma \theta$.
For that purpose we will compute the first variation of this action in
$\left(\mathfrak{p}^*\otimes\Lambda^8T^*\mathcal{P}\right)
\oplus_\mathcal{P}\left(\mathfrak{p}\otimes T^*\mathcal{P}\right)$
and write under which condition on a submanifold $\Gamma$ this first variation of $\mathcal{A}$ vanishes for all
variations of $\Gamma$ which respect (\ref{contrainteVariationnelle}).

First because of the independence condition (\ref{independence}) we can always assume that,
locally, $\Gamma$ is a graph over $\mathcal{P}$ or, in other words, the image of a section
$\varphi$ of the bundle $\left(\mathfrak{p}^*\otimes\Lambda^8T^*\mathcal{P}\right)
\oplus_\mathcal{P}\left(\mathfrak{p}\otimes T^*\mathcal{P}\right)$ over $\mathcal{P}$.
Thus we can write $\mathcal{A}[\Gamma] = \int_\mathcal{P}\varphi^*\theta$ and we can coordinatize
an infinitesimal variation of $\Gamma$ by maps on $\mathcal{P}$ $\delta \eta$ and
$\delta \psi$ with compact supports.
The first variation of $\mathcal{A}$ can then be written:
\[
 \delta\mathcal{A}_\Gamma(\delta \eta,\delta \psi) = \int_\mathcal{P} 
\delta \psi_A\wedge \varphi^*\left(d\eta^A +\frac{1}{2}[\eta\wedge \eta]^A\right)
+ (\varphi^*\psi_A)\wedge \left(d(\delta\eta^A) + [\delta\eta\wedge \varphi^*\eta]^A\right).
\]
We note that
\[
 (\varphi^*\psi_A)\wedge d(\delta\eta^A) 
= d(\delta\eta^A\wedge \varphi^*\psi_A) + \delta\eta^A\wedge \varphi^*d\psi_A
\]
and
\[
 (\varphi^*\psi_A)\wedge [\delta\eta\wedge \varphi^*\eta]^A
= (\varphi^*\psi_A)\wedge c^A_{BC}\delta\eta^B\wedge \varphi^*\eta^C
 = -\delta\eta^B\wedge \varphi^*(\hbox{ad}_\eta^*\wedge \psi)_B,
\]
where $(\hbox{ad}_\eta^*\wedge \psi)_B:= c^A_{CB}\eta^C\wedge \psi_A$
(see (\ref{defiadxi*})). Thus, assuming that
$(\delta \eta,\delta \psi)$
has a compact support,
\begin{equation}\label{variationPremiere}
 \delta\mathcal{A}_\Gamma(\delta \eta,\delta \psi) = \int_\mathcal{P} 
 \delta \psi_A\wedge \varphi^*\left(d\eta^A +\frac{1}{2}[\eta\wedge \eta]^A\right)
+ \delta\eta^A\wedge \varphi^*(d\psi - \hbox{ad}_\eta^*\wedge \psi)_A.
\end{equation}
Solutions to the HVDW equations are the submanifolds $\Gamma$ which satisfy the constraints
(\ref{contrainteVariationnelle}) and which are such that $\delta\mathcal{A}_\Gamma(\delta \eta,\delta \psi)$
vanishes for any infinitesimal variations 
$(\delta \eta,\delta \psi) = (\delta \alpha,\delta \omega,\delta \psi)$ which respect this constraint,
i.e. which satisfy
\begin{equation}\label{contrainteVariationnelleLinearisee}
 \delta \alpha^c\wedge \alpha^d\wedge \varphi^*\psi_A
+ \alpha^c\wedge \delta \alpha^d\wedge \varphi^*\psi_A
+ \alpha^c\wedge \alpha^d\wedge \delta \psi_A 
 = \kappa_A^{cd}\delta\eta^B\wedge \varphi^*\eta_B^{(9)}.
\end{equation}
In other words the solutions are characterized by the fact that Condition (\ref{contrainteVariationnelleLinearisee})
implies the following
\begin{equation}\label{tovanish}
 \delta \psi_A\wedge \varphi^*\left(d\eta^A +\frac{1}{2}[\eta\wedge \eta]^A\right)
+ \delta\eta^A\wedge \varphi^*(d\psi - \hbox{ad}_\eta^*\wedge \psi)_A = 0.
\end{equation}

\subsection{Parametrization of infinitesimal variations satisfying (\ref{contrainteVariationnelleLinearisee})}\label{section42}
Let us set 
\[
 \delta \eta^A = \lambda^A_a\alpha^a +  \lambda^A_i\omega^i,
\]
\[
 \delta\psi_A = \frac{1}{2}\chi{_A}^{cd}\alpha^{(2)}_{cd}\wedge \omega^{(6)}
- \chi{_A}^{ck}\alpha^{(3)}_c\wedge \omega^{(5)}_k
+ \frac{1}{2}\chi{_A}^{jk}\alpha^{(4)}\wedge \omega^{(4)}_{jk}
\]
where $\lambda^A_C,\chi{_A}^{CD}$ are smooth function with compact support on $\mathcal{P}$,
and define $\varpi:= \varphi^*\psi$ and
\begin{equation}\label{rappeldecomposition}
\begin{array}{ccl}
 \varpi_A:= \varphi^*\psi_A & = & \displaystyle \frac{1}{2}\kappa_A^{cd}\alpha^{(2)}_{cd}\wedge \omega^{(6)}
- (\psi{_A}^{ck}\circ \varphi)\alpha^{(3)}_c\wedge \omega^{(5)}_k
+ \frac{1}{2}(\psi{_A}^{jk}\circ \varphi)\alpha^{(4)}\wedge \omega^{(4)}_{jk}\\
& = & \displaystyle \frac{1}{2}\kappa_A^{cd}\alpha^{(2)}_{cd}\wedge \omega^{(6)}
- \varpi{_A}^{ck}\alpha^{(3)}_c\wedge \omega^{(5)}_k
+ \frac{1}{2}\varpi{_A}^{jk}\alpha^{(4)}\wedge \omega^{(4)}_{jk}
\end{array}
\end{equation}
Thus we may write (\ref{contrainteVariationnelleLinearisee}) as:
\[
 (\lambda^c_{c'}\kappa_A^{c'd} - \lambda^c_k\varpi{_A}^{dk})
+ (\lambda^d_{d'}\kappa_A^{cd'} + \lambda^d_k\varpi{_A}^{ck})
+ \chi{_A}^{cd} = (\lambda^b_b+\lambda^i_i)\kappa_A^{cd}.
\]
Hence we can express $\chi{_A}^{cd}$ in terms of the other quantities:
\[
 \chi{_A}^{cd} = \lambda^b_{b'}(\delta^{b'}_b\kappa_A^{cd} - \delta^c_b\kappa_A^{b'd}- \delta^d_b\kappa_A^{cb'})
+ \lambda^b_k(\delta^c_b\varpi{_A}^{dk} - \delta^d_b\varpi{_A}^{ck}) + \lambda^i_i\kappa_A^{cd}.
\]
Thus (\ref{contrainteVariationnelleLinearisee}) means that we can express $\delta\psi$ in terms of
$\lambda^A_B$, $\chi{_A}^{ck}$ and $\chi{_A}^{jk}$:
\[
 \begin{array}{ccr}
  \delta\psi_A & = & \frac{1}{2}\left[\lambda^b_{b'}\left(\delta^{b'}_b\kappa_A^{cd}\alpha^{(2)}_{cd}
  - \kappa_A^{b'd}\alpha^{(2)}_{bd} - \kappa_A^{cb'}\alpha^{(2)}_{cb}\right) \right. \hfill\\
& & \left.  + \lambda^b_k(\varpi{_A}^{dk}\alpha^{(2)}_{bd} - \varpi{_A}^{ck}\alpha^{(2)}_{cb})
+ \lambda^i_i\kappa_A^{cd}\alpha^{(2)}_{cd}\right] \wedge \omega^{(6)}\\
  & & - \chi{_A}^{ck}\alpha^{(3)}_c\wedge \omega^{(5)}_k
+ \frac{1}{2}\chi{_A}^{jk}\alpha^{(4)}\wedge \omega^{(4)}_{jk}\\
  & = & \left[\lambda^b_{b'}\left(\frac{1}{2}\delta^{b'}_b\kappa_A^{cd}\alpha^{(2)}_{cd}
  -  \kappa_A^{b'd}\alpha^{(2)}_{bd} \right)
  + \lambda^b_k\varpi{_A}^{dk}\alpha^{(2)}_{bd}
+ \frac{1}{2}\lambda^i_i\kappa_A^{cd}\alpha^{(2)}_{cd}\right] \wedge \omega^{(6)}\\
  & & - \chi{_A}^{ck}\alpha^{(3)}_c\wedge \omega^{(5)}_k
+ \frac{1}{2}\chi{_A}^{jk}\alpha^{(4)}\wedge \omega^{(4)}_{jk}.
 \end{array}
\]
\subsection{The Euler--Lagrange equations}\label{section43}
On the one hand, setting $\Omega:= \varphi^*\left(d\eta +\frac{1}{2}[\eta\wedge \eta]\right)$,
we can decompose
\begin{equation}\label{decompostionOmega}
 \Omega^A = \frac{1}{2}Q{^A}_{cd}\alpha^c\wedge \alpha^d
+ Q{^A}_{ck}\alpha^c\wedge \omega^k + \frac{1}{2}Q{^A}_{jk}\omega^j\wedge \omega^k,
\end{equation}
so that, taking into account (\ref{contrainteVariationnelleLinearisee}),
the first term on the l.h.s. of (\ref{tovanish}) reads:
\[
\begin{array}{ccl}
\delta\psi_A\wedge\Omega^A & = & \displaystyle
\left[\lambda^b_{b'}\left(\frac{1}{2}\delta^{b'}_b\kappa_A^{cd}Q{^A}_{cd}
  - \kappa_A^{b'd}Q{^A}_{bd} \right)
  + \lambda^b_k\varpi{_A}^{dk}Q{^A}_{bd} + \frac{1}{2}\lambda^i_i\kappa_A^{cd}Q{^A}_{cd} \right.\\
  & & 
 \displaystyle \hfill \left. + \chi{_A}^{ck}Q{^A}_{ck} + \chi{_A}^{jk}Q{^A}_{jk}\right]\eta^{(10)}.
\end{array}
\]
On the other hand, setting $̀\nabla^\eta \varpi:= \varphi^*(d\psi - \hbox{ad}_\eta^*\wedge \psi)$
for short, and
using the decomposition ̀$(\nabla^\eta \varpi)_A = (\nabla^\eta \varpi)_A^a\alpha^{(3)}_a \wedge \omega^{(6)}
+ (\nabla^\eta \varpi)_A^i\alpha^{(4)} \wedge \omega^{(5)}_i$, the second term in the l.h.s. of (\ref{tovanish})
 taking into account (\ref{contrainteVariationnelleLinearisee}) reads
\[
 \delta\eta^A\wedge (\nabla^\eta \varpi)_A = \left(\lambda^A_a(\nabla^\eta \varpi)^a_A
+ \lambda^A_i(\nabla^\eta \varpi)^i_A\right)\eta^{(10)}.
\]
In conclusion [(\ref{contrainteVariationnelleLinearisee}) $\Longrightarrow$ (\ref{tovanish})] is equivalent to
the condition that
\[
 \begin{array}{ccccl}
  \lambda^b_{b'}\left(\frac{1}{2}\delta^{b'}_b\kappa_A^{cd}Q{^A}_{cd}
  - \kappa_A^{b'd}Q{^A}_{bd} + (\nabla^\eta \varpi)^{b'}_b\right) & + & 
  \lambda^b_k\left(\varpi{_A}^{dk}Q{^A}_{bd} + (\nabla^\eta \varpi)^k_b\right) &  & \\ 
  + \lambda^j_{b'}(\nabla^\eta \varpi)^{b'}_j & + &
  \lambda^j_k\left(\frac{1}{2}\delta^k_j\kappa_A^{cd}Q{^A}_{cd} + (\nabla^\eta \varpi)^k_j\right) & & \\
  + \chi{_A}^{ck}Q{^A}_{ck} & + & \chi{_A}^{jk}Q{^A}_{jk} & = & 0
 \end{array}
\]
be satisfied
for all $\lambda^b_{b'},\lambda^b_k,\lambda^j_{b'},\lambda^j_k,\chi{_A}^{ck}$ and $\chi{_A}^{jk}$.

Hence the HVDW equations or, equivalentely, the Euler--Lagrange equations of the action
$\int_\Gamma \theta$, are
\begin{eqnarray}
 (\nabla^\eta \varpi)^{b'}_b & = &  \kappa_A^{b'd}Q{^A}_{bd}
- \frac{1}{2}\delta^{b'}_b\kappa_A^{cd}Q{^A}_{cd}\label{EL1}\\
(\nabla^\eta \varpi)^k_b & = & - \varpi{_A}^{dk}Q{^A}_{bd}\label{EL2}\\
(\nabla^\eta \varpi)^{b'}_j & = & 0\label{EL3}\\
(\nabla^\eta \varpi)^k_j  & = & - \frac{1}{2}\delta^k_j\kappa_A^{cd}Q{^A}_{cd}\label{EL4}\\
Q{^A}_{ck}  & = & 0\label{EL5}\\
 Q{^A}_{jk} & = & 0\label{EL6}
\end{eqnarray}

\section{Study of the solutions of the HVDW equations}

The first four equations (\ref{EL1})
to (\ref{EL4}) can be translated into the following relations on
$(\nabla^\eta \varpi)_A = (\nabla^\eta \varpi)^a_A\alpha_a^{(3)}\wedge \omega^{(4)}
+ (\nabla^\eta \varpi)^i_A\alpha^{(4)}\wedge \omega^{(5)}_i$ for $A = a$ or
$j$:
\begin{equation}\label{preEL1a4}
 \left\{
\begin{array}{ccccc}(\nabla^\eta \varpi)_a & = & (\kappa^{bc}_AQ{^A}_{ac} - \hbox{S}\delta^b_a)
\alpha^{(3)}_b\wedge \omega^{(6)}
& - & \varpi{_A}^{cj}Q{^A}_{ac} \alpha^{(4)}\wedge \omega^{(5)}_j\\
 (\nabla^\eta \varpi)_j & = & & - & \hbox{S}\alpha^{(4)}\wedge \omega^{(5)}_j,
\end{array}
\right.
\end{equation}
where
\[
 \hbox{S}:= \frac{1}{2}\kappa^{cd}_AQ{^A}_{cd}.
\]
Alternatively we can also introduce coefficients $u^{ia}_b$ (see, in the Appendix,
(\ref{a86}), (\ref{a87}) and (\ref{a88}))
and replace $(\nabla^\eta \varpi)_j$ by:
\[
 (\nabla^\eta \varpi){_a}^b:= (\nabla^\eta \varpi)_ju^{jb}_a.
\]
(then $(\nabla^\eta \varpi)_j = \frac{1}{2}(\nabla^\eta \varpi){_a}^bu^a_{jb}$). Then
equations (\ref{preEL1a4}) are equivalent to
\begin{equation}\label{EL1a4}
\left\{
\begin{array}{ccccc}(\nabla^\eta \varpi)_a & = & (\kappa^{bc}_AQ{^A}_{ac} - S\delta^b_a)
\alpha^{(3)}_b\wedge \omega^{(6)}
& - & \varpi{_A}^{cj}Q{^A}_{ac} \alpha^{(4)}\wedge \omega^{(5)}_j\\
 (\nabla^\eta \varpi){_a}^b & = & & - & \hbox{S}u^{jb}_a\alpha^{(4)}\wedge \omega^{(5)}_j,
\end{array}
\right.
\end{equation}
On the other hand, by using (\ref{decompostionOmega}), we see that Equations (\ref{EL5}) and (\ref{EL6})
are equivalent to:
\begin{equation}\label{decompostionOmegaAfter}
 \varphi^*(d\eta+\frac{1}{2}[\eta\wedge \eta])^A = \frac{1}{2}Q{^A}_{cd}\alpha^c\wedge \alpha^d,
\end{equation}
or equivalentely
\begin{eqnarray}
 (d\alpha + \omega\wedge \alpha)^a & = & \frac{1}{2}Q{^a}_{cd}\alpha^c\wedge \alpha^d
\label{decompostionOmegaAfteralpha} \\
(d\omega + \omega\wedge \omega)^i & = & \frac{1}{2}Q{^i}_{cd}\alpha^c\wedge \alpha^d
\label{decompostionOmegaAfteromega}
\end{eqnarray}
In the following we first exploit Equations (\ref{decompostionOmegaAfteralpha}) and
(\ref{decompostionOmegaAfteromega}). Then we analyze the content of Equation 
(\ref{EL1a4}).

\subsection{The spontaneous fibration lemma}\label{paragraphfibration}

\begin{lemm}\label{spontaneousfibration}
 Let $\eta = (\alpha,\omega)$ be a 1-form defined on 10-dimensional manifold $\mathcal{P}$ 
 with coefficients in $\mathfrak{p}$. Assume that the rank of $\eta$ is maximal, equal to
10 everywhere and that there exist functions $Q{^A_{}}{^{}_{bc}}$ on $\mathcal{P}$
 such that (\ref{decompostionOmegaAfteralpha}) and (\ref{decompostionOmegaAfteromega}) are satisfied.

Then, for any point $\textsf{m}$ of $\mathcal{P}$, there exists a neighborhood $\mathcal{P}_\textsf{m}$
of $\textsf{m}$ on which there exist local coordinate
functions $(x,g)=(x^0,x^1,x^2,x^3,g)$ with values in $\R^4\times \mathfrak{G}$, such that
 \begin{equation}\label{trivialiseAlpha}
  \alpha^a = (g^{-1})^a_{a'}e^{a'},\quad \hbox{where }e^{a'} = e^{a'}_\mu(x)dx^\mu 
 \end{equation}
and
\begin{equation}\label{trivialiseOmega}
 \omega{^a_{}}{_b^{}} = (g^{-1})^a_{a'}A^{a'}_{b'}g^{b'}_b + (g^{-1})^a_{a'}dg^{a'}_b,
\quad \hbox{where }A^{a'}_{b'} = A^{a'}_{b'\mu}(x)dx^\mu .
\end{equation}
As a consequence the set $\mathcal{X}_\textsf{m}$ of submanifolds of $\mathcal{P}_\textsf{m}$
of equation $x=$ constant has a structure of 4-dimensional manifold and the quotient map
$\pi:= \mathcal{P}_\textsf{m}\longrightarrow \mathcal{X}_\textsf{m}$ is a local fibration.
Moreover $\alpha$ and $\omega$ are the lifts on the total space of his local fibre bundle
of respectively a solder form and a connection form of a pseudo-Riemannian structure on
$\mathcal{X}_\textsf{m}$.
\end{lemm}
\emph{Proof} --- \textbf{Step 1} --- Consider the Pfaffian system
\begin{equation}\label{pfaffien1}
 \alpha^a|_\textsf{f} = 0,\quad \forall a= 0,1,2,3,
\end{equation}
where the unknown $\textsf{f}$ is a 6-dimensional submanifold of $\mathcal{P}$.
Because of (\ref{decompostionOmegaAfteralpha}) we have:
\[
 d\alpha^a = \left( - \omega{^a_{}}{_b^{}} + \frac{1}{2}Q{^a_{}}{^{}_{cb}}\alpha^c\right) \wedge \alpha^b,
\]
which means that the Pfaffian system (\ref{pfaffien1}) is integrable and satisfies the hypotheses
of Frobenius' theorem. By applying this theorem we deduce that through any point $\textsf{m}\in 
\mathcal{P}$ there exists a unique 6-dimensional submanifold $\textsf{f}$ which is a solution of the system
(\ref{pfaffien1}). This defines a fibration
$\pi_\textsf{m}:\mathcal{P}_\textsf{m}\longrightarrow \mathcal{X}_\textsf{m}$
of a neighborhood $\mathcal{P}_\textsf{m}$ of $\textsf{m}$ in $\mathcal{P}$
with values in a neighborhood $\mathcal{X}_\textsf{m}$ of the space of leaves which
are solutions of (\ref{pfaffien1}). We choose local coordinates
$x^0\cdots,x^3$ on $\mathcal{X}_\textsf{m}$.
Abusing notation we will set $x^\mu \simeq x^\mu\circ \pi_\textsf{m}$.
We also choose 6 extra local coordinate functions $y^1,\cdots y^6$ on a neighborhood of
$\textsf{m}$ (which we still call $\mathcal{P}_\textsf{m}$) such that the submanifolds
of equation $y^\mu =$ constant, $\forall \mu=1,\cdots ,6$ are transverse to the leaves
$\textsf{f}$. Hence we can assume without loss of generality that the 10 functions
$x^0,\cdots,x^3,y^1,\cdots y^6$ form a system of local coordinates on $\mathcal{P}_\textsf{m}$.\\
\textbf{Step 2} --- Let us denote by $\Sigma$ the submanifold of equation
$y^1= \cdots = y^6 = 0$. 
%
%
We deduce from (\ref{decompostionOmega}) and (\ref{decompostionOmegaAfteromega}) that
\begin{equation}\label{pfaffienPre2}
 d\omega{^a_{}}{_b^{}} + \omega{^a_{}}{_{a'}{}} \wedge \omega{^{a'}_{}}{_b^{}} = \frac{1}{2}Q{^a_{}}{^{}_{bcd}}
 \alpha^c\wedge \alpha^d
\end{equation}
and hence, in particular, by restriction to a leaf $\textsf{f}$:
\begin{equation}\label{pfaffien2}
 \left(d\omega{^a_{}}{_b^{}} + \omega{^a_{}}{_{a'}{}} \wedge \omega{^{a'}_{}}{_b^{}}\right)|_\textsf{f} = 0.
\end{equation}
This means that the Pfaffian system in $\textsf{f}\times \mathfrak{G}$
\begin{equation}\label{pfaffien3}
 (dg - g\omega)|_\textsf{f} = 0
\end{equation}
is integrable and, in particular, there exists a unique solution which is equal to
$1_\mathfrak{G}$ at the intersection point of $\textsf{f}$ and $\Sigma$.
We hence obtain a map $g:\mathcal{P}_\textsf{m}\longrightarrow \mathfrak{G}$
which is equal to $1_\mathfrak{G}$ on $\Sigma$ and which satisfies (\ref{pfaffien3}).
Since the family $\left(\omega{^a_{}}{_b^{}}|_\textsf{f}\right)_{0\leq a<b\leq 3}$ 
form a coframe on $\textsf{f}$, we deduce from (\ref{pfaffien3}) that the components 
$\gamma^4|_\textsf{f},\cdots,\gamma^9|_\textsf{f}$ in a basis $\mathfrak{g}$ of the restriction of
$\gamma:= g^{-1}dg$ to $\textsf{f}$ form also a coframe on $\textsf{f}$.\\
\textbf{Step 3} ---Relation (\ref{pfaffien3}) also means that $\omega - g^{-1}dg$ is a linear combination
of the forms $\alpha^0,\cdots,\alpha^3$ or equivalentely of the forms $dx^0,\cdots,dx^3$.
Thus there exist real valued functions $A^a_{\mu b}$ of $x$ and $g$,
for $0\leq \mu\leq 3$, or,
equivalentely, functions $A_\mu$ with values in $\mathfrak{g}$ such that
\[
 \omega = g^{-1}dg + g^{-1}A_\mu(x,g) gdx^\mu.
\]
But then $d\omega+\omega\wedge\omega = g^{-1}(dA+A\wedge A)g$ and
$\omega$ satisfies (\ref{pfaffienPre2}) iff $A_\mu$
does not depend on $g$, i.e.
\begin{equation}
 \omega = g^{-1}dg + g^{-1}A_\mu(x) gdx^\mu
\end{equation}
or (\ref{trivialiseOmega}). Similarly if we set $\alpha:= g^{-1}e_\mu(x,g)dx^\mu$, we get
$d\alpha +\omega \wedge \alpha = g^{-1}(de+A\wedge e)$. Hence the relation
\[
 d\alpha^a + \omega{^a_{}}{_b^{}} \wedge \alpha^b =
 \frac{1}{2}Q{^a_{}}{^{}_{bc}}\alpha^b\wedge \alpha^c
\]
implies that $e_\mu$ does not depend on $g$, thus (\ref{trivialiseAlpha}) follows.\hfill $\square$\\

\subsection{Change of unknown functions}\label{paragraphlong}
To summarize the result of the previous section we can build local coordinate $(x,g)$, where $x\in \R^4$ and
$g\in \mathfrak{G}$ and we can write
\begin{equation}\label{rappelalphaeomegaA}
 \alpha^a = (g^{-1})^a_{a'}e^{a'}
\quad \hbox{and}\quad
\omega{^a}_b = (g^{-1})^a_{a'}dg^{a'}_b + (g^{-1})^a_{a'}A{^{a'}}_{b'}g^{b'}_b,
\end{equation}
where $e^a$ and $A{^a}_b$ are 1-forms which depends \emph{only} on the $x$ variables.
Equivalentely,
\[
 (\alpha,\omega) = (0,g^{-1}dg) + \hbox{Ad}_{g^{-1}}H,
\]
where $H = (e,A)$ is a $\mathfrak{p}$-valued 1-form whose coefficients depend only on the
$x$ variables. For analyzing Equations (\ref{EL1a4}) it will be useful to express them
using coordinates $(x,g)$ and functions adapted to these coordinates.

\subsubsection{Replacing the 8-forms $\varpi$}
We replace the 8-forms $\varpi$ defined in Section \ref{section42} by
\begin{equation}\label{changepsitop}
 p:= \hbox{Ad}_{g^{-1}}^*\varpi
\end{equation}
and we set:
\begin{equation}\label{nablap}
\nabla^H p:= dp - \hbox{ad}_H^*\wedge p.
\end{equation}
By using (\ref{deficoadjoint}) in the Appendix this definition reads
\begin{eqnarray}
 (\nabla^H p)_a & = & dp_a - p_b\wedge A{^b}_a\label{nablapa}\\
(\nabla^H p){_a}^b & = & dp{_a}^b + A{^b}_c\wedge p{_a}^c - p{_c}^b\wedge A{^c}_a + 2p_a\wedge e^b.\label{nablapab}
\end{eqnarray}
Recall (Section \ref{section43}) that $\nabla^\eta \varpi = \varphi^*(d\psi - \hbox{ad}_\eta^*\wedge \psi)
= d\varpi - \hbox{ad}_{(\alpha,\omega)}^*\wedge \varpi$. It follows from
(\ref{lastcoro}) that
\begin{equation}\label{changedpsitodp}
\nabla^H p =  \hbox{Ad}_{g^{-1}}^*(\nabla^\eta \varpi).
\end{equation}
This means that $(\nabla^H p){_a} = (g^{-1})^{a'}_a(\nabla^\eta \varpi){_{a'}}$ and
$(\nabla^H p){_a}^b = (g^{-1})^{a'}_ag^b_{b'}(\nabla^\eta \varpi){_{a'}}^{b'}$.
Hence (\ref{EL1a4}) translates as
\begin{equation}\label{EL1a41}
\left\{
\begin{array}{ccccc}
 (\nabla^H p)_a & = & (\kappa^{bc}_A(g^{-1})^{a'}_aQ{^A}_{a'c} - \hbox{S}(g^{-1})^b_a)
\alpha^{(3)}_b\wedge \omega^{(6)}
& - & \varpi{_A}^{cj}(g^{-1})^{a'}_aQ{^A}_{a'c} \alpha^{(4)}\wedge \omega^{(5)}_j\\
 (\nabla^H p){_a}^b & = & & - & \hbox{S}(g^{-1})^{a'}_ag^b_{b'}u^{jb'}_{a'}\alpha^{(4)}\wedge \omega^{(5)}_j
\end{array}
\right.
\end{equation}

\subsubsection{Replacing coefficients $Q{^A}_{cd}$}
Let us define the tensors $T{^a}_{cd}$ (torsion) and $R{^a}_{bcd}$ (Riemann curvature) such that 
\begin{equation}\label{definitionTetR}
(de + A\wedge e)^a =  \frac{1}{2}T{^a}_{cd}e^c\wedge e^d
\quad\hbox{and}\quad
(dA + A\wedge A){^a}_b:= \frac{1}{2}R{^a}_{bcd}e^c\wedge e^d,
\end{equation}
which clearly depend only on $x$ (and not on $g$).
Using (\ref{rappelalphaeomegaA}) we compute that
$(d\alpha + \omega\wedge \alpha)^a = (g^{-1})^a_{a'}(de+A\wedge e)^{a'}$ and
$(d\omega + \omega \wedge \omega){^a}_b = (g^{-1})^a_{a'}g^{b'}_b(dA+A\wedge A){^{a'}}_{b'}$.
Hence, by using (\ref{decompostionOmegaAfteralpha}) and (\ref{rappelalphaeomegaA}), we find that
\[
\begin{array}{rcccl}
\displaystyle \frac{1}{2}Q{^a}_{cd}\alpha^c\wedge \alpha^d 
& = & \displaystyle \frac{1}{2}(g^{-1})^a_{a'}T{^{a'}}_{c'd'}e^{c'}\wedge e^{d'}
& = & \displaystyle \frac{1}{2}(g^{-1})^a_{a'}T{^{a'}}_{c'd'}g^{c'}_cg^{d'}_d \alpha^c\wedge \alpha^d\\
\displaystyle \frac{1}{2}Q{^a}_{bcd}\alpha^c\wedge \alpha^d 
& = & \displaystyle \frac{1}{2}(g^{-1})^a_{a'}g^{b'}_bR{^{a'}}_{b'c'd'}e^{c'}\wedge e^{d'}
& = & \displaystyle \frac{1}{2}(g^{-1})^a_{a'}g^{b'}_bR{^{a'}}_{b'c'd'}g^{c'}_cg^{d'}_d \alpha^c\wedge \alpha^d.
\end{array}
\]
Thus
\begin{equation}\label{QetT}
 Q{^a}_{cd} = (g^{-1})^a_{a'}g^{c'}_cg^{d'}_d T{^{a'}}_{c'd'},
\end{equation}
\begin{equation}\label{QetR}
 Q{^a}_{bcd} = (g^{-1})^a_{a'}g^{b'}_bg^{c'}_cg^{d'}_d R{^{a'}}_{b'c'd'}.
\end{equation}
Now consider the following term, which appears in the r.h.s. of (\ref{EL1a41}):
\[
 \kappa^{bc}_A(g^{-1})^{a'}_aQ{^A}_{a'c} = 2u_i^{bc}(g^{-1})^{a'}_aQ{^i}_{a'c}
= 2u_{ic'}^b\textsf{h}^{c'c}(g^{-1})^{a'}_a Q{^i}_{a'c}
= 2\textsf{h}^{c'c}(g^{-1})^{a'}_a Q{^b}_{c'a'c},
\]
it follows from (\ref{QetR}) that
\[
 \kappa^{bc}_A(g^{-1})^{a'}_aQ{^A}_{a'c}
= 2\textsf{h}^{c'c}(g^{-1})^{a'}_a (g^{-1})^b_{b'}g_{c'}^{c''}g_{a'}^{a''}g_{c}^{d}
  R{^{b'}}_{c''a''d} = 2\textsf{h}^{c''d} (g^{-1})^b_{b'} R{^{b'}}_{c''ad},
\]
where we used $\textsf{h}^{c'c}g_{c'}^{c''}g_{c}^{d} = \textsf{h}^{c''d}$.
Thus, by posing $R{^{ab}}_{cd}:= \textsf{h}^{bb'} R{^a}_{b'cd}$, we obtain
that
\[
 \kappa^{bc}_A(g^{-1})^{a'}_aQ{^A}_{a'c} = 2(g^{-1})^b_{b'}R{^{b'd}}_{ad}.
\]
We recognize the Ricci tensor:
set $\hbox{Ric}{^b}_a:= R{^{bd}}_{ad}$, then the previous relation reads
\begin{equation}\label{simplificationkQ}
 \kappa^{bc}_A(g^{-1})^{a'}_aQ{^A}_{a'c} = 2(g^{-1})^b_{b'} \hbox{Ric}{^{b'}}_a.
\end{equation}
We can also express the quantity $\hbox{S} = \frac{1}{2}\kappa^{ac}_AQ{^A}_{ac}$: (\ref{simplificationkQ})
is equivalent to $\kappa^{bc}_AQ{^A}_{ac} = 2g^{a'}_a(g^{-1})^b_{b'} \hbox{Ric}{^{b'}}_{a'}$
hence
\begin{equation}\label{simplificationtrkQ}
 \hbox{S} = \hbox{Ric}{^a}_a,
\end{equation}
which is nothing but the scalar curvature.
Lastly using again (\ref{QetR}) and (\ref{QetT}) we have
\[
\begin{array}{ccl}
 \varpi{_A}^{cj}(g^{-1})^{a'}_aQ{^A}_{a'c} & = & (g^{-1})^{a'}_a\left(
\frac{1}{2}\varpi{_d}^{bcj}Q{^d}_{ba'c} + \varpi{_d}^{cj}Q{^d}_{a'c}\right)\\
& = & (g^{-1})^{d}_{d'}g^{c'}_c\left(\frac{1}{2}\varpi{_d}^{bcj}g^{b'}_bR{^{d'}}_{b'ac'}
+ \varpi{_d}^{cj}T{^{d'}}_{ac'}\right)
\end{array}
\]
Using (\ref{simplificationkQ}) and the previous relation we transform
the first equation of (\ref{EL1a41}) into
\[
\begin{array}{ccr}
 (\nabla^H p)_a  & = & (2(g^{-1})^b_{b'} \hbox{Ric}{^{b'}}_a - \hbox{S}(g^{-1})^b_a)
\alpha^{(3)}_b\wedge \omega^{(6)}\hfill \\
& & \hfill -  (g^{-1})^{d}_{d'}g^{c'}_c\left(\frac{1}{2}\varpi{_d}^{bcj}g^{b'}_bR{^{d'}}_{b'ac'}
+ \varpi{_d}^{cj}T{^{d'}}_{ac'}\right)
\alpha^{(4)}\wedge \omega^{(5)}_j
\end{array}
\]
Thus introducing the \emph{Einstein tensor}
\[
 \hbox{E}{^b}_a := \hbox{Ric}{^b}_a - \frac{1}{2}\hbox{S}\delta^b_a
\]
and observing that $(g^{-1})^{a'}_ag^b_{b'}u^{jb'}_{a'} = (\hbox{Ad}_{g^{-1}}^*u^j){_a}^b$
(see (\ref{a89})) we can write (\ref{EL1a41}) as:
\begin{equation}\label{EL1a42}
\left\{
\begin{array}{ccc}
 (\nabla^H p)_a & = & 2(g^{-1})^b_{b'}\hbox{E}{^{b'}}_a \alpha^{(3)}_b\wedge \omega^{(6)}\hfill \\
 & & \hfill - (g^{-1})^{d}_{d'}g^{c'}_c\left(\frac{1}{2}\varpi{_d}^{bcj}g^{b'}_bR{^{d'}}_{b'ac'}
+ \varpi{_d}^{cj}T{^{d'}}_{ac'}\right)
\alpha^{(4)}\wedge \omega^{(5)}_j\\
 (\nabla^H p){_a}^b & = & \hfill  - \hbox{S}(\hbox{Ad}_{g^{-1}}^*u^j){_a}^b \alpha^{(4)}\wedge \omega^{(5)}_j
\end{array}
\right.
\end{equation}

\subsubsection{Replacing the forms $(\alpha,\omega)$}
The previous equations give the decomposition of the 9-form $\nabla^H p$ in the basis
$(\alpha^{(3)}_a\wedge \omega^{(6)},\alpha^{(4)}\wedge \omega^{(5)}_i)$. Let $e^a$
be the forms defined by (\ref{rappelalphaeomegaA}) and let $\gamma = \gamma^iu_i := g^{-1}dg$.
We want to use the coframe $(e^0,\cdots,e^3,\gamma^4,\cdots,\gamma^9)$ and
to replace $\alpha^{(3)}_a\wedge \omega^{(6)} = \frac{\partial}{\partial \alpha^a}\iN \alpha^{(4)}\wedge \omega^{(6)}$
and $\alpha^{(4)}\wedge \omega^{(5)}_i = \frac{\partial}{\partial \omega^i}\iN \alpha^{(4)}\wedge \omega^{(6)}$
in terms of $e^{(3)}_a\wedge \gamma^{(6)}:= \frac{\partial}{\partial e^a}\iN e^{(4)}\wedge \gamma^{(6)}$
and $e^{(4)}\wedge \gamma^{(5)}_i:= \frac{\partial}{\partial \gamma^i}\iN e^{(4)}\wedge \gamma^{(6)}$
(see Section \ref{notations} for the notations). For that
it suffices to note that $e^{(4)}\wedge \gamma^{(6)} = \eta^{(10)} = \alpha^{(4)}\wedge \omega^{(6)}$
(because in particular $\omega = \gamma + Ad_{g^{-1}}A$) and to use the relations
\begin{equation}\label{passageDomegaaDgamma}
\left\{
 \begin{array}{ccl}
  \frac{\partial}{\partial \alpha^a} & = & g_a^{a'}\left(\frac{\partial}{\partial e^{a'}}
  - (\hbox{Ad}_{g^{-1}}A_{a'})^i\frac{\partial}{\partial \gamma^i}\right),\\
  \frac{\partial}{\partial \omega^i} & = & \frac{\partial}{\partial \gamma^i}
 \end{array}
 \right.
\end{equation}
where $(\hbox{Ad}_{g^{-1}}A_{a'})^i:= u^i(\hbox{Ad}_{g^{-1}}A_{a'})$. Hence
\[
\left\{
 \begin{array}{ccl}
 \alpha^{(3)}_a\wedge \omega^{(6)} = \frac{\partial}{\partial \alpha^a}\iN \eta^{(10)}
 & = & g_a^{a'}\left(e^{(3)}_{a'}\wedge \gamma^{(6)}
- (\hbox{Ad}_{g^{-1}}A_{a'})^ie^{(4)}\wedge \gamma^{(5)}_i\right)\\
 \alpha^{(4)}\wedge \omega^{(5)}_i = \frac{\partial}{\partial \omega^i}\iN \eta^{(10)}
 & = & e^{(4)}\wedge \gamma^{(5)}_i.
 \end{array}
 \right.
\]
Thus substituting these expressions in the r.h.s. of (\ref{EL1a42}) we obtain
\begin{equation}\label{EL1a43}
\left\{
\begin{array}{ccc}
 (\nabla^H p)_a & = & 2\hbox{E}{^b}_a e^{(3)}_b\wedge \gamma^{(6)} \hfill - 2\hbox{E}{^b}_a
 (\hbox{Ad}_{g^{-1}}A_b)^je^{(4)}\wedge \gamma^{(5)}_j  \hfill \\
 & & \hfill - (g^{-1})^{d}_{d'}g^{c'}_c\left(\frac{1}{2}\varpi{_d}^{bcj}g^{b'}_bR{^{d'}}_{b'ac'}
+ \varpi{_d}^{cj}T{^{d'}}_{ac'}\right)
e^{(4)}\wedge \gamma^{(5)}_j\\
 (\nabla^H p){_a}^b & = & \hfill  - \hbox{S}(\hbox{Ad}_{g^{-1}}^*u^j){_a}^b e^{(4)}\wedge \gamma^{(5)}_j
\end{array}
\right.
\end{equation}

\subsubsection{Replacing all the components of $\varpi$}
We need to go further and also to compute
\[
\begin{array}{ccl}
 \alpha^{(2)}_{cd}\wedge \omega^{(6)} & = & \frac{\partial}{\partial \alpha^d}\iN \alpha^{(3)}_c\wedge \omega^{(6)}
 = \frac{\partial}{\partial \alpha^d}\iN
 g_c^{c'}\left(e^{(3)}_{c'}\wedge \gamma^{(6)} - (\hbox{Ad}_{g^{-1}}A_{c'})^ie^{(4)}\wedge \gamma^{(5)}_i\right)\\
 & = & g_d^{d'}\left(\frac{\partial}{\partial e^{d'}}
  - (\hbox{Ad}_{g^{-1}}A_{d'})^j\frac{\partial}{\partial \gamma^j}\right) \iN
  g_c^{c'}\left(e^{(3)}_{c'}\wedge \gamma^{(6)} - (\hbox{Ad}_{g^{-1}}A_{c'})^ie^{(4)}\wedge \gamma^{(5)}_i\right)\\
  & = & g_c^{c'}g_d^{d'}\left( e^{(2)}_{c'd'}\wedge \gamma^{(6)}
+ (\hbox{Ad}_{g^{-1}}A_{d'})^je^{(3)}_{c'}\wedge \gamma^{(5)}_j
  - (\hbox{Ad}_{g^{-1}}A_{c'})^ie^{(3)}_{d'}\wedge \gamma^{(5)}_i
  \right.\\
  & & \hfill \left. + (\hbox{Ad}_{g^{-1}} A_{c'})^i(\hbox{Ad}_{g^{-1}}A_{d'})^je^{(4)}\wedge \gamma^{(4)}_{ij}\right)\\
  & = & g_c^{c'}g_d^{d'}\left( e^{(2)}_{c'd'}\wedge \gamma^{(6)}
 + \left((\hbox{Ad}_{g^{-1}}A_{d'})^ie^{(3)}_{c'} - (\hbox{Ad}_{g^{-1}}A_{c'})^ie^{(3)}_{d'}\right)\wedge \gamma^{(5)}_i
 \right.\\
  & & \hfill \left. 
  + (\hbox{Ad}_{g^{-1}}A_{c'})^i(\hbox{Ad}_{g^{-1}}A_{d'})^je^{(4)}\wedge \gamma^{(4)}_{ij}\right),
\end{array}
\]
second
\[
\begin{array}{ccl}
 \alpha^{(3)}_c\wedge \omega^{(5)}_j & = & \frac{\partial}{\partial \alpha^c}\iN \alpha^{(4)}\wedge \omega^{(5)}_j
 = \frac{\partial}{\partial \alpha^c}\iN e^{(4)}\wedge \gamma^{(5)}_j\\
 & = & g_c^{c'}\left(\frac{\partial}{\partial e^{c'}}
  -(\hbox{Ad}_{g^{-1}} A_{c'})^k\frac{\partial}{\partial \gamma^k}\right)\iN e^{(4)}\wedge \gamma^{(5)}_j\\
  & = & g_c^{c'}\left( e^{(3)}_{c'}\wedge \gamma^{(5)}_j - (\hbox{Ad}_{g^{-1}}A_{c'})^k
  e^{(4)}\wedge \gamma^{(4)}_{jk}\right)
\end{array}
\]
and lastly
\[
\begin{array}{ccl}
 \alpha^{(4)}\wedge \omega^{(4)}_{jk} & = & \frac{\partial}{\partial \gamma^k}\iN \alpha^{(4)}\wedge \omega^{(5)}_j
 = \frac{\partial}{\partial \gamma^k}\iN e^{(4)}\wedge \gamma^{(5)}_j\\
 & = & e^{(4)}\wedge \gamma^{(4)}_{jk}.
\end{array}
\]
Now we can relate two decompositions of $\varpi_A$. On the one hand, starting from (\ref{rappeldecomposition}):
\[
 \begin{array}{ccr}
  \varpi_A & = & \frac{1}{2}\varpi{_A}^{cd}\alpha^{(2)}_{cd}\wedge \omega^{(6)}
- \varpi{_A}^{ck}\alpha^{(3)}_c\wedge \omega^{(5)}_k
+ \frac{1}{2}\varpi{_A}^{jk}\alpha^{(4)}\wedge \omega^{(4)}_{jk}\hfill \\
& = & \frac{1}{2}\varpi{_A}^{cd}g_c^{c'}g_d^{d'}\left( e^{(2)}_{c'd'}\wedge \gamma^{(6)}
  + \left((\hbox{Ad}_{g^{-1}}A_{d'})^je^{(3)}_{c'} - (\hbox{Ad}_{g^{-1}}A_{c'})^je^{(3)}_{d'}\right)\wedge \gamma^{(5)}_j
  \right. \hfill \\ 
& & \hfill \left. 
  + (\hbox{Ad}_{g^{-1}}A_{c'})^j(\hbox{Ad}_{g^{-1}}A_{d'})^k e^{(4)}\wedge \gamma^{(4)}_{jk}\right)\\
& &  - \varpi{_A}^{cj}g_c^{c'}
  \left( e^{(3)}_{c'}\wedge \gamma^{(5)}_j - (\hbox{Ad}_{g^{-1}}A_{c'})^k e^{(4)}\wedge \gamma^{(4)}_{jk}\right)
  + \frac{1}{2}\varpi{_A}^{jk}e^{(4)}\wedge \gamma^{(4)}_{jk}\\
 & = & \frac{1}{2}\varpi{_A}^{cd}g_c^{c'}g_d^{d'}e^{(2)}_{c'd'}\wedge \gamma^{(6)}
  + \left(\varpi{_A}^{cd}g_c^{c'}g_d^{d'}(\hbox{Ad}_{g^{-1}}A_{d'})^j  - \varpi{_A}^{cj}g_c^{c'} \right)
  e^{(3)}_{c'}\wedge \gamma^{(5)}_j 
  \hfill \\
& &  \hfill + \left( \frac{1}{2}\varpi{_A}^{cd}g_c^{c'}g_d^{d'}
(\hbox{Ad}_{g^{-1}}A_{c'})^j(\hbox{Ad}_{g^{-1}}l_{d'})^k
  + \varpi{_A}^{cj}g_c^{c'}(\hbox{Ad}_{g^{-1}}A_{c'})^k
+ \frac{1}{2}\varpi{_A}^{jk}\right)e^{(4)}\wedge \gamma^{(4)}_{jk}.
 \end{array}
\]
On the other hand if we decompose $p_A =  \frac{1}{2}p{_A}^{cd}e^{(2)}_{cd}\wedge \gamma^{(6)}
- p{_A}^{ck}e^{(3)}_c\wedge \gamma^{(5)}_k + \frac{1}{2}p{_A}^{jk}e^{(4)}\wedge \gamma^{(4)}_{jk}$
and we develop the relation $\varpi = \hbox{Ad}_g^*p$, we get
\[
  \varpi_A = (\hbox{Ad}_g^*p)_A = \frac{1}{2}(\hbox{Ad}_g^*p){_A}^{cd}e^{(2)}_{cd}\wedge \gamma^{(6)}
- (\hbox{Ad}_g^*p){_A}^{ck}e^{(3)}_c\wedge \gamma^{(5)}_k+ \frac{1}{2}(\hbox{Ad}_g^*p){_A}^{jk}e^{(4)}\wedge \gamma^{(4)}_{jk}.
\]
By identification we deduce the following
\begin{equation}\label{identificationcd}
 (\hbox{Ad}_g^*p){_A}^{cd} = \varpi{_A}^{c'd'}g_{c'}^cg_{d'}^d, 
\end{equation}
and $(\hbox{Ad}_g^*p){_A}^{cj} = \varpi{_A}^{c'j}g_{c'}^{c}
 - \varpi{_A}^{c'd'}g_{c'}^{c}g_{d'}^{d} (\hbox{Ad}_{g^{-1}}A_{d})^j$, from which
we deduce by using (\ref{identificationcd})
\begin{equation}\label{identificationck}
 (\hbox{Ad}_g^*p){_A}^{cj} 
 = \varpi{_A}^{c'j}g_{c'}^{c} - (\hbox{Ad}_g^*p){_A}^{cd}(\hbox{Ad}_{g^{-1}}A_{d})^j.
\end{equation}
We could also derive a relation between $p{_A}^{cd}$ and $\varpi{_A}^{cd}$ which we will not 
write since we don't neeed it.
Relation (\ref{identificationcd}) is equivalent to
\[
 p{_A}^{cd} = (\hbox{Ad}_{g^{-1}}^*\varpi){_A}^{c'd'}g_{c'}^cg_{d'}^d.
\]
It gives us for $p{_a}^{bcd} := u^{ib}_a p{_i}^{cd}$:
\[
 \begin{array}{ccl}
 p{_a}^{bcd} & = & (g^{-1})_a^{a'}g^b_{b'}\varpi{_{a'}}^{b'c'd'}g_{c'}^cg_{d'}^d
= (g^{-1})_a^{a'}g^b_{b'}g_{c'}^cg_{d'}^d \kappa{_{a'}}^{b'c'd'} \\
 & = & (g^{-1})_a^{a'}g^b_{b'}g_{c'}^cg_{d'}^d
(\delta^{c'}_{a'}\textsf{h}^{b'd'} - \delta^{d'}_{a'}\textsf{h}^{b'c'})\\
 & = & \delta^{c}_{a}\textsf{h}^{bd} - \delta^{d}_{a}\textsf{h}^{bc} = \kappa{_{a}}^{bcd};
\end{array}
\]
and for $p{_a}^{cd}$:
$p{_a}^{cd} = (g^{-1})_a^{a'}\varpi{_{a'}}^{c'd'}g_{c'}^cg_{d'}^d 
 = (g^{-1})_a^{a'}g_{c'}^cg_{d'}^d\kappa{_{a}}^{cd} = 0$.
Hence we deduce that the coefficients of $p$ satisfy
\begin{equation}\label{tetedep}
p{_a}^{cd} = 0
\quad\hbox{and}\quad
 p{_a}^{bcd} = \kappa{_{a}}^{bcd}.
\end{equation}
Moreover Relation (\ref{identificationck}) is equivalent to
\begin{equation}\label{identicationck1}
 \varpi{_A}^{cj} = (g^{-1})^c_{c'}(\hbox{Ad}_g^*p){_A}^{c'j}
+ (g^{-1})^c_{c'}(\hbox{Ad}_g^*p){_A}^{c'd}(\hbox{Ad}_{g^{-1}}A_d)^j
\end{equation}
and give us for $\varpi{_A}^{cj} = \varpi{_a}^{bcj}$:
\[
 \varpi{_a}^{bcj} = (g^{-1})^c_{c'} g_a^{a'}(g^{-1})^b_{b'}p{_{a'}}^{b'c'j}
+ (g^{-1})^c_{c'}g_a^{a'}(g^{-1})^b_{b'}p{_{a'}}^{b'c'd}(\hbox{Ad}_{g^{-1}}A_d)^j
\]
and thus by using (\ref{tetedep})
\begin{equation}\label{dernierlhs1}
 (g^{-1})^{a'}_ag^b_{b'}g^c_{c'}\varpi{_{a'}}^{b'c'j} = p{_a}^{bcj} 
 + \kappa{_a}^{bcd}(\hbox{Ad}_{g^{-1}}A_d)^j.
\end{equation}
Similarly (\ref{identicationck1}) gives us for $\varpi{_A}^{cj} = \varpi{_a}^{cj}$:
\[
 \varpi{_a}^{cj} = (g^{-1})^c_{c'} g_a^{a'}p{_{a'}}^{c'j}
+ (g^{-1})^c_{c'}g_a^{a'}p{_{a'}}^{c'd}(\hbox{Ad}_{g^{-1}}A_d)^j
\]
and hence by using (\ref{tetedep})
\begin{equation}\label{dernierlhs2}
 (g^{-1})^{a'}_{a}g^{c}_{c'}\varpi{_{a'}}^{c'j} = p{_a}^{cj}.
\end{equation}
We now use Relations (\ref{dernierlhs1}) and (\ref{dernierlhs2}) for eliminating
$\varpi{_d}^{cj}$ and $\varpi{_d}^{bcj}$ in the r.h.s. of (\ref{EL1a43})
and write
\[
  (g^{-1})^d_{d'}g^{c'}_c\left(\frac{1}{2}g_b^{b'}\varpi{_d}^{bcj}R{^{d'}}_{b'ac'}
+ \varpi{_d}^{cj}T{^{d'}}_{ac'}\right) 
 = \frac{1}{2}(p{_d}^{bcj} + \kappa{_d}^{bce}(\hbox{Ad}_{g^{-1}}A_e)^j)R{^d}_{bac}
 + p{_d}^{cj}T{^d}_{ac}
\]
But since $\kappa{_d}^{bce}R{^d}_{bac} = (\delta^c_d\textsf{h}^{be}
- \delta^e_d\textsf{h}^{bc})R{^d}_{bac} = - 2\hbox{Ric}{^e}_a$,
\[
 (g^{-1})^d_{d'}g^{c'}_c\left(\frac{1}{2}g_b^{b'}\varpi{_d}^{bcj}R{^{d'}}_{b'ac'}
+ \varpi{_d}^{cj}T{^{d'}}_{ac'}\right) 
 = - \hbox{Ric}{^b}_a(\hbox{Ad}_{g^{-1}}A_b)^j
- \frac{1}{2}p{_d}^{bcj}R{^d}_{bca} - p{_d}^{cj}T{^d}_{ca}.
\]
Hence we can write (\ref{EL1a43}) as
\begin{equation}\label{EL1a44}
\left\{
\begin{array}{ccr}
 (\nabla^H p)_a & = & 2\hbox{E}{^b}_a e^{(3)}_b\wedge \gamma^{(6)} \hfill - 2\hbox{E}{^b}_a
 (\hbox{Ad}_{g^{-1}}A_b)^je^{(4)}\wedge \gamma^{(5)}_j   \\
 & & \hfill + \left(\frac{1}{2}p{_d}^{bcj}R{^d}_{bca} + p{_d}^{cj}T{^d}_{ca}
+ \hbox{Ric}{^b}_a(\hbox{Ad}_{g^{-1}}A_b)^j \right)
e^{(4)}\wedge \gamma^{(5)}_j\\
 (\nabla^H p){_a}^b & = & \hfill  - \hbox{S}(\hbox{Ad}_{g^{-1}}^*u^j){_a}^b e^{(4)}\wedge \gamma^{(5)}_j
\end{array}
\right.
\end{equation}

\subsubsection{The left hand side}
We first prove a preliminary lemma.
\begin{lemm}\label{preliminarylemma}
 Let $\Gamma^a_{bc}$ (Christoffel symbols) be the functions depending on $x$ such
that $A{^a}_c = \Gamma^a_{bc}e^b$. Then
\begin{equation}\label{formuledea}
 de^a = \left(\frac{1}{2}T{^a}_{c'd'} - \Gamma^a_{c'd'}\right)e^{c'}\wedge e^{d'}
\end{equation}
and, as a consequence,
\begin{equation}\label{formuledee3}
 de^{(3)}_c = Y_ce^{(4)},
\end{equation}
where $Y_c:= T{^d}_{cd} - \Gamma^d_{cd}+\Gamma^d_{dc}$.
\end{lemm}
\emph{Proof} --- By (\ref{definitionTetR}) we have 
$de^a + A{^a}_{d'}\wedge e^{d'} = \frac{1}{2}T{^a}_{c'd'}e^{c'}\wedge e^{d'}$, hence
by substituting $A{^a}_{d'} = \Gamma^a_{c'd'}e^{c'}$, we obtain (\ref{formuledea}).
Then we compute 
\[
 de^{(3)}_c = de^d\wedge e^{(2)}_{cd} =
\left(\frac{1}{2}T{^d}_{c'd'} - \Gamma^d_{c'd'}\right)e^{c'}\wedge e^{d'} \wedge e^{(2)}_{cd}
\]
from which (\ref{formuledee3}) follows.\hfill $\square$\\

In the previous section we have collected the algebraic constraints
which have to be imposed in $p$, namely Relations (\ref{tetedep}).
It remains to compute the l.h.s. of (\ref{EL1a44}) taking into account these constraints.
We start from the decomposition:
\[
 p_A =  \frac{1}{2}p{_A}^{cd}e^{(2)}_{cd}\wedge \gamma^{(6)}
- p{_A}^{ck}e^{(3)}_c\wedge \gamma^{(5)}_k + \frac{1}{2}p{_A}^{jk}e^{(4)}\wedge \gamma^{(4)}_{jk}
\]
which, taking into account (\ref{tetedep}), reads equivalentely as
\begin{equation}\label{paapres}
 p_a =  0 - p{_a}^{ck}e^{(3)}_c\wedge \gamma^{(5)}_k + \frac{1}{2}p{_a}^{jk}e^{(4)}\wedge \gamma^{(4)}_{jk}
\end{equation}
and, using $\kappa{_a}^{bcd}e^{(2)}_{cd} = 2\textsf{h}^{bc}e^{(2)}_{ac}$,
\begin{equation}\label{pabapres}
 p{_a}^b = \textsf{h}^{bc}e^{(2)}_{ac}\wedge \gamma^{(6)}
- p{_a}^{bck}e^{(3)}_c\wedge \gamma^{(5)}_k + \frac{1}{2}p{_a}^{bjk}e^{(4)}\wedge \gamma^{(4)}_{jk}.
\end{equation}
Using (\ref{nablapa}) and (\ref{paapres}) we get
\[
 \begin{array}{ccl}
  (\nabla^Hp)_a & = & -dp{_a}^{ck}\wedge e^{(3)}_c\wedge \gamma^{(5)}_k
- p{_a}^{ck}de^{(3)}_c\wedge \gamma^{(5)}_k
+ p{_a}^{ck}e^{(3)}_c\wedge d\gamma^{(5)}_k\\
& & + \frac{1}{2}dp{_a}^{jk}\wedge e^{(4)}\wedge \gamma^{(4)}_{jk}
+ \frac{1}{2}p{_a}^{jk}de^{(4)}\wedge \gamma^{(4)}_{jk}
+ \frac{1}{2}p{_a}^{jk}e^{(4)}\wedge d\gamma^{(4)}_{jk}\\
& & -\left(-p{_b}^{ck}e^{(3)}_c\wedge \gamma^{(5)}_k 
 + \frac{1}{2}p{_b}^{jk}e^{(4)}\wedge \gamma^{(4)}_{jk}\right)
\wedge \Gamma^b_{c'a}e^{c'}.
 \end{array}
\]
Hence using Lemmas \ref{preliminarylemma} and \ref{lemma2} and using the notation
$df = f_{;a}e^a + f_{;i}\gamma^i$ for any function $f$, this gives us
\[
 \begin{array}{ccl}
  (\nabla^Hp)_a & = &  - p{_a}{^{ck}}_{;c} e^{(4)}\wedge \gamma^{(5)}_k
 + p{_a}{^{ck}}_{;k} e^{(3)}_c\wedge \gamma^{(6)}
- p{_a}^{ck}Y_c e^{(4)}\wedge \gamma^{(5)}_k\\
& & + p{_a}{^{jk}}_{;k} e^{(4)}\wedge \gamma^{(5)}_{j}
- \frac{1}{2}p{_a}^{jk}e^{(4)}\wedge c^l_{jk}\gamma^{(5)}_l\\
& & + p{_b}^{ck}\Gamma^b_{ca}e^{(4)}\wedge \gamma^{(5)}_{k}.
 \end{array}
\]
Thus
\begin{equation}\label{finallhsa}
 \begin{array}{ccl}
 (\nabla^Hp)_a & = & \left( -  p{_a}{^{cj}}_{;c}- p{_a}^{cj}Y_c+ p{_b}^{cj}\Gamma^b_{ca}
+ p{_a}{^{jk}}_{;k} - \frac{1}{2}p{_a}^{kl}c^j_{kl}\right)e^{(4)}\wedge \gamma^{(5)}_{j}\\
& & + p{_a}{^{bk}}_{;k} e^{(3)}_b\wedge \gamma^{(6)}.
 \end{array}
\end{equation}
We now turn to the computation of $(\nabla^Hp){_a}^b$ (using  (\ref{nablapab})).
As a preliminary, consider 
$q$, the $\mathfrak{p}^*$-valued 2-form such that $q_A = \kappa_A^{dc}e_{cd}^{(2)}$
(hence $(q_A) = (q_a,q{_a}^b)$ with $q_a = 0$ and $q{_a}^b:= \textsf{h}^{bc}e_{ac}^{(2)}$),
and compute
$(\nabla^H q){_a}^b = dq{_a}^b + A{^b}_c\wedge q{_a}^c - q_{c}^b\wedge A{^c}_a + 2q_a\wedge e^b$:
by using $de_{ac}^{(2)} = de^d\wedge e_{acd}^{1)}$, we get
\[
 \begin{array}{ccl}
  (\nabla^H q){_a}^b  & = & 
\textsf{h}^{bc}\left( T{^d}_{cd}e_a^{(3)} +  T{^d}_{da}e_c^{(3)} +  T{^d}_{ac}e_d^{(3)} 
- A{^d}_d\wedge e_{ac}^{(2)} - A{^d}_c\wedge e_{da}^{(2)} - A{^d}_a\wedge e_{cd}^{(2)} \right)\\
& & \hfill + \textsf{h}^{cd} A{^b}_c\wedge e_{ad}^{(2)} - \textsf{h}^{bd} A{^c}_a\wedge e_{cd}^{(2)}
 \end{array}
\]
Setting $A^{ab}:= \textsf{h}^{bb'}A{^a}_{b'}$ and noting that $A^{ab} + A^{ba} = 0$
and $A{^d}_d = 0$, we have
\[
 \begin{array}{ccl}
  (\nabla^H q){_a}^b  & = & 
\textsf{h}^{bc}\left( T{^d}_{cd}e_a^{(3)} +  T{^d}_{da}e_c^{(3)} +  T{^d}_{ac}e_d^{(3)} \right)
- \textsf{h}^{bc}A{^d}_d\wedge e_{ac}^{(2)} - A^{db}\wedge e_{da}^{(2)}
- \textsf{h}^{bc}A{^d}_a\wedge e_{cd}^{(2)} \\
& & \hfill +  A^{bd}\wedge e_{ad}^{(2)} - \textsf{h}^{bc} A{^d}_a\wedge e_{dc}^{(2)}\\
& = & \textsf{h}^{bc}\left( T{^d}_{cd}e_a^{(3)} +  T{^d}_{da}e_c^{(3)} +  T{^d}_{ac}e_d^{(3)} \right)\\
& = & \left(\textsf{h}^{bd}T{^c}_{ad} - \textsf{h}^{bc}T{^d}_{ad} + \textsf{h}^{be}T{^d}_{ed} \delta^c_a\right)e_c^{(3)}.
 \end{array}
\]
Thus
\begin{equation}\label{partiel}
 (\nabla^H q\wedge \gamma^{(6)}){_a}^b =
\left(\textsf{h}^{bd}T{^c}_{ad} - \textsf{h}^{bc}T{^d}_{ad} + \textsf{h}^{be}T{^d}_{ed} \delta^c_a\right)
e_c^{(3)}\wedge \gamma^{(6)}.
\end{equation}
Note that $q\wedge \gamma^{(6)}$ is the `first part' of $p$, i.e. the component which is
a multiple of $\gamma^{(6)}$. It remains to compute the other part, i.e. 
$(\nabla^H \overline{p}){_a}^b$, where $\overline{p}:= p - q\wedge \gamma^{(6)}$. This computation is
similar to the one for $(\nabla^Hp)_a$.
\[
 \begin{array}{ccl}
  (\nabla^H \overline{p}){_a}^b & = & -dp{_a}^{bck}\wedge e^{(3)}_c\wedge \gamma^{(5)}_k
- p{_a}^{bck}de^{(3)}_c\wedge \gamma^{(5)}_k
+ p{_a}^{bck}e^{(3)}_c\wedge d\gamma^{(5)}_k\\
& & + \frac{1}{2}dp{_a}^{bjk}\wedge e^{(4)}\wedge \gamma^{(4)}_{jk}
+ \frac{1}{2}p{_a}^{bjk}de^{(4)}\wedge \gamma^{(4)}_{jk}
+ \frac{1}{2}p{_a}^{bjk}e^{(4)}\wedge d\gamma^{(4)}_{jk}\\
& & + \Gamma^{b}_{c'a'}e^{c'}\wedge \left(-p{_{a}}^{a'ck}e^{(3)}_c\wedge \gamma^{(5)}_k
+ \frac{1}{2}p{_a}^{a'jk}e^{(4)}\wedge d\gamma^{(4)}_{jk} \right)\\
& & -\left(-p{_{b'}}^{bck}e^{(3)}_c\wedge \gamma^{(5)}_k 
+ \frac{1}{2}p{_{b'}}^{bjk}e^{(4)}\wedge d\gamma^{(4)}_{jk}
\right) \wedge \Gamma^{b'}_{c'a}e^{c'}\\
& & \hfill
+2 \left(-p{_{a}}^{ck}e^{(3)}_c\wedge \gamma^{(5)}_k
+ \frac{1}{2}p{_a}^{jk}e^{(4)}\wedge d\gamma^{(4)}_{jk} \right)\wedge e^b
 \end{array}
\]
Hence as before
\[
 \begin{array}{ccl}
  (\nabla^H \overline{p}){_a}^b & = &  
- p{_a}^{bck}{_{;c}} e^{(4)}\wedge \gamma^{(5)}_k
+ p{_a}{^{bck}}{_{;k}} e^{(3)}_c\wedge \gamma^{(6)}
- p{_a}^{bck}Y_c e^{(4)}\wedge \gamma^{(5)}_k\\
& & + p{_a}^{bjk}{_{;k}} e^{(4)}\wedge \gamma^{(5)}_{j}
- \frac{1}{2}p{_a}^{bjk}c^i_{jk}e^{(4)}\wedge \gamma^{(5)}_{i}\\
& & - \Gamma^{b}_{ca'} p{_{a}}^{a'ck}e^{(4)}\wedge \gamma^{(5)}_k
+ \Gamma^{b'}_{ca}p{_{b'}}^{bck}e^{(4)}\wedge \gamma^{(5)}_k  
- 2 p{_{a}}^{bk}e^{(4)}\wedge \gamma^{(5)}_k
 \end{array}
\]
Thus
\[
 \begin{array}{ccl}
  (\nabla^H \overline{p}){_a}^b & = &  \left(
- p{_a}^{bcj}{_{;c}} - p{_a}^{bcj}Y_c
- \Gamma^{b}_{ca'} p{_{a}}^{a'cj} 
+ \Gamma^{b'}_{ca}p{_{b'}}^{bcj} - 2 p{_{a}}^{bj}\right.\\
& & \hfill \left.+ p{_a}^{bjk}{_{;k}}  - \frac{1}{2}p{_a}^{bkl}c^j_{kl} \right)
e^{(4)}\wedge \gamma^{(5)}_j\\
& &  + p{_a}{^{bck}}{_{;k}} e^{(3)}_c\wedge \gamma^{(6)}
 \end{array}
\]
and, using (\ref{partiel}) and $p = q\wedge \gamma^{(6)} + \overline{p}$,
\begin{equation}\label{finallhsab}
 \begin{array}{ccl}
  (\nabla^H p){_a}^b & = & 
  \left(\textsf{h}^{bd}T{^c}_{ad} - \textsf{h}^{bc}T{^d}_{ad} + \textsf{h}^{be}T{^d}_{ed} \delta^c_a
  + p{_a}{^{bck}}{_{;k}}\right) e_c^{(3)}\wedge \gamma^{(6)} \\
&&  + \left(
- p{_a}^{bcj}{_{;c}} - p{_a}^{bcj}Y_c
- \Gamma^{b}_{ca'} p{_{a}}^{a'cj} 
+ \Gamma^{b'}_{ca}p{_{b'}}^{bcj} - 2 p{_{a}}^{bj}\right.\\
& & \hfill \left.+ p{_a}^{bjk}{_{;k}}  - \frac{1}{2}p{_a}^{bkl}c^j_{kl} \right)
e^{(4)}\wedge \gamma^{(5)}_j
 \end{array}
\end{equation}

\subsubsection{Conclusion: the HVDW equations}

We now can write the dynamical equations completely in terms of the fields
$A$, $e$ and $p$. We identify the l.h.s. of (\ref{EL1a44}) by using formulas
(\ref{finallhsa}) and (\ref{finallhsab}).
This gives us for the component of $(\nabla^Hp){_a}$ along $e_{b}^{(3)}\wedge \gamma^{(6)}$:
\begin{equation}\label{brutELab}
 p{_a}{^{bk}}_{;k} = 2\hbox{E}{^b}_a ,\quad \forall a,b,
\end{equation}
for the component of $(\nabla^Hp){_a}$ along $e^{(4)}\wedge \gamma^{(5)}_j$:
\begin{equation}\label{brutELaj}
 \begin{array}{rcl}
 - p{_a}{^{cj}}_{;c}- p{_a}^{cj}Y_c+ p{_b}^{cj}\Gamma^b_{ca} & & \\
  + p{_a}{^{jk}}_{;k} - \frac{1}{2}p{_a}^{kl}c^j_{kl} & = & 
  -(2\hbox{E}{^b}_a - \hbox{Ric}{^b}_a) (\hbox{Ad}_{g^{-1}}A_b)^j\\
& & + \frac{1}{2}p{_d}^{bcj}R{^d}_{bca} + p{_d}^{cj}T{^d}_{ca},\quad \forall a,j,
 \end{array}
\end{equation}
for the component of $(\nabla^Hp){_a}^b$ along $e_{(c)}^{(3)}\wedge \gamma^{(6)}$:
\begin{equation}\label{brutELabc}
 \textsf{h}^{bd}T{^c}_{ad} - \textsf{h}^{bc}T{^d}_{ad} + \textsf{h}^{be}T{^d}_{ed} \delta^c_a
  + p{_a}{^{bck}}{_{;k}} = 0,\quad \forall a,b,c,
\end{equation}
and for the component of $(\nabla^Hp){_a}^b$ along $e^{(4)}\wedge \gamma^{(5)}_j$:
\begin{equation}\label{brutELabj}
 \begin{array}{rcl}
 - p{_a}^{bcj}{_{;c}} - p{_a}^{bcj}Y_c
- \Gamma^{b}_{ca'} p{_{a}}^{a'cj} 
+ \Gamma^{b'}_{ca}p{_{b'}}^{bcj} - 2 p{_{a}}^{bj} & & \\
+ p{_a}^{bjk}{_{;k}}  - \frac{1}{2}p{_a}^{bkl}c^j_{kl} &
= & - \hbox{S}(\hbox{Ad}_{g^{-1}}^*u^j){_a}^b,\quad \forall a,b,j.
 \end{array}
\end{equation}
By using the fact that Relation (\ref{brutELabc}) implies
$T{^a}_{ca} = - \frac{1}{2}\textsf{h}_{cd}p{_a}{^{daj}}{_{;j}}$ one can see that
(\ref{brutELabc}) is equivalent to:
\begin{equation}\label{brutELabcbis}
 T{^a}_{cd} = - \left( \textsf{h}_{de}\delta^a_{a'}\delta^{c'}_c 
 + \frac{1}{2}\delta^{c'}_{a'}(\delta^a_d\textsf{h}_{ce} - \delta^a_c\textsf{h}_{de})\right)
 p{_{c'}}{^{ea'j}}_{;j}.
\end{equation}
We can organize these equations into two systems
\begin{equation}\label{firstsystem}
\left\{
 \begin{array}{lcl}
  \hbox{E}{^b}_a & = & \frac{1}{2} p{_a}{^{bj}}_{;j}\\
   T{^a}_{cd} & = & - \left( \textsf{h}_{de}\delta^a_{a'}\delta^{c'}_c 
 + \frac{1}{2}\delta^{c'}_{a'}(\delta^a_d\textsf{h}_{ce}
 - \delta^a_c\textsf{h}_{de})\right) p{_{c'}}{^{ea'j}}_{;j}
 \end{array}
 \right.
\end{equation}
and
\begin{equation}\label{secondsystem}
\left\{
 \begin{array}{rcl}
   p{_a}{^{cj}}_{;c} + p{_a}^{cj}Y_c - p{_b}^{cj}\Gamma^b_{ca}
   + \frac{1}{2}p{_d}^{bcj}R{^d}_{bca} + p{_d}^{cj}T{^d}_{ca} & & \\
 \hfill   -(2\hbox{E}{^b}_a - \hbox{Ric}{^b}_a) (\hbox{Ad}_{g^{-1}}A_b)^j
  & = &  p{_a}{^{jk}}_{;k} - \frac{1}{2}p{_a}^{kl}c^j_{kl}\\
   p{_a}^{bcj}{_{;c}} + p{_a}^{bcj}Y_c
+ \Gamma^{b}_{ca'} p{_{a}}^{a'cj} 
- \Gamma^{b'}_{ca}p{_{b'}}^{bcj} + 2 p{_{a}}^{bj}& & \\
\hfill - \hbox{S}(\hbox{Ad}_{g^{-1}}^*u^j){_a}^b
& = &  p{_a}^{bjk}{_{;k}}  - \frac{1}{2}p{_a}^{bkl}c^j_{kl} 
 \end{array}
 \right.
\end{equation}

\section{Consequences of the equations}\label{sectiondiscussion}
\subsection{Global results}
We first remark that, if a basis $(\mathfrak{l}_A)_A$ of $\mathfrak{p}$ is fixed,
we can associate to any $\mathfrak{p}$-valued 1-form $(\alpha,\omega)$ which is of rank 10
everywhere the Riemannian
metric $G:= (\alpha^0)^2+\cdots +(\alpha^3)^2+(\omega^4)^2+\cdots +(\omega^9)^2$
on $\mathcal{P}$. In the relativistic case this metric depends on the choice of the basis 
$(\mathfrak{l}_A)_A$ and should not have any physical meaning in general. Nevertheless it
has the virtue of being always
positive definite and hence, in any case, it defines a topology on $\mathcal{P}$ which does not depend on
the choice of the basis $(\mathfrak{l}_A)_A$.

\begin{prop}\label{propositionglobale}
Assume that $\mathfrak{G}$ is simply connected (i.e. it is the Spin group).
 Let $(\varpi,\alpha, \omega)$ be a solution of the HVDW equations and assume that the
$\mathfrak{p}$-valued 1-form $(\alpha,\omega)$ is of rank 10 everywhere. Assume that
$\mathcal{P}$ endowed with the topology induced by the metric $G$ as above is complete, connected and
open. Then any leaf $\textsf{f}$ is a diffeomorphic to a quotient of $\mathfrak{G}$
by a group action.
\end{prop}
\emph{Proof} --- Since $\eta$ is of rank
10 everywhere we can construct a family of tangent vector fields $(\xi_4,\cdots,\xi_9)$ on $\mathcal{P}$
defined by $\alpha^a(\xi_i) = 0$, $\forall a,i$ and $\omega^j(\xi_i) = \delta^j_i$,
$\forall i,j$.
We can interpret Equation (\ref{pfaffien3}) as the simultaneous flow equations of these
vector fields. Then (\ref{pfaffien2}) means that these vector fields are in involution.
These vector fields are obviously uniformly bounded in the topology induced by $G$, hence they
are complete, since $\mathcal{P}$ is complete. Hence we can integrate them for all time and get
a covering map from $\mathfrak{G}$ to the leaf $\textsf{f}$.\hfill $\square$\\

In the Riemannian case $\mathfrak{G}$ is compact. Proposition \ref{propositionglobale}
has then further consequences.
\begin{coro}\label{corocompact}
 Assume that $(\vec{\mathbb{M}},\textsf{h})$ is the Euclidean space and the same hypotheses
of Proposition \ref{propositionglobale}.
Then $\mathcal{P}$ is the total space of a principal bundle over a 4-dimensional manifold
with fibers diffeomorphic to $Spin(4)$ or $SO(4)$.
\end{coro}
\emph{Proof} --- We apply the previous Proposition: each leaf has $Spin(4)$ as a universal
cover, hence is diffeomorphic to $Spin(4)$ or $SO(4)$. But these leaves are also
compact, which allows us to apply a result of  Ehresmann \cite{Ehresmann50} to conclude.\hfill $\square$\\

\subsection{The Riemannian case}
\begin{theo}\label{theoremEuclide}
Assume that $(\vec{\mathbb{M}},\textsf{h})$ is the Euclidean space and that
$\mathfrak{G}$ is simply connected (i.e. it is the Spin group).
Let $(\varpi,\alpha, \omega)$ be a solution of the HVDW equations and assume that the
$\mathfrak{p}$-valued 1-form $(\alpha,\omega)$ is of rank 10 everywhere. Assume that
$\mathcal{P}$ endowed with the topology induced by the metric $G$ as above is complete, connected and
open. Then $\mathcal{P}$ is the total space of a principal bundle over a 4-dimensional manifold
$\mathcal{X}$ with fibers diffeomorphic to $Spin(4)$ or $SO(4)$.
Moreover $\omega$ defines the Levi-Civita connection associated to the metric on $\mathcal{X}$ defined by
$\alpha$ and $\mathcal{X}$ is an Einstein manifold.
\end{theo}
\emph{Proof} --- We first apply Corollary \ref{corocompact}. Then
the proof follows the same lines as in \cite{helein14} for Yang--Mills fields.
We know that the left hand sides of (\ref{firstsystem}) does not depend on the variables
$g$ but only on $x$. Hence the same is true for the right hand sides, e.g. for
$p{_a}{^{bj}}_{;j}$. Let $\textsf{f}$ be a fiber over the point $\textsf{x}\in \mathcal{X}$.
We observe that $p{_a}{^{bj}}_{;j}\gamma^{(6)}|_\textsf{f} = d(p{_a}{^{bj}}\gamma^{(5)}_j)|_\textsf{f}$.
Since $\textsf{f}$ is compact without boundary, we have
\[
 p{_a}{^{bj}}_{;j}\int_\textsf{f}\gamma^{(6)} = \int_\textsf{f} p{_a}{^{bj}}_{;j}\gamma^{(6)}
=  \int_\textsf{f}d(p{_a}{^{bj}}\gamma^{(5)}_j) = 0.
\]
A similar reasoning gives $p{_{c'}}{^{ea'j}}_{;j} = 0$. Hence the right hand sides of (\ref{firstsystem})
vanish, which implies the conclusion.\hfill $\square$

\subsection{The relativistic case: a discussion}\label{paragraphdiscussion}
If $(\vec{\mathbb{M}},\textsf{h})$ is the Minkowski space, the situation is much more complicated, because
the structure group is not compact. 

First there is no analogue of Corollary \ref{corocompact} in general and we could not
exclude a priori
complete, connected solutions $(\mathcal{P},\alpha,\omega,\varpi)$
for which the leaves of the foliation are dense and thus the quotient space would not
be separated. We will not discuss such solutions, since they are far from the standard
definition of a space-time in General Relativity.
However they could lead to interesting models in the framework of non-commutative geometry.

Note that, beside the metric $G$ constructed on $\mathcal{P}$ in the previous section,
we could also privilege non degenerate bilinear forms on a solution 
$(\mathcal{P},\alpha,\omega,\varpi)$ of the HVWD equations of the type
$K:= \textsf{h}_{ab}\alpha^a\alpha^b + \textsf{K}_{ij}\omega^i\omega^j$, where
$\textsf{K}_{ij}$ is a non degenerate bilinear form on $\mathfrak{g}$ which is
invariant by the adjoint action of $\mathfrak{G}$. Such forms are not positive
definite in general, but they do not depend on the choice of a basis of $\mathfrak{p}$
and they may possibly have a physical sense\footnote{For instance in the degenerate
case where $\textsf{K}= 0$, if $(\alpha,\omega)$ is a solution
of the HVDW equations, then $K$ is locally the pull-back by the fibration map of the pseudo-Riemannian
metric on the quotient space of leaves found in Lemma \ref{spontaneousfibration}.}.
Understanding the geometry of the quotient space of leaves in this framework seems even more
difficult a priori, but it is perhaps more relevant from a physical point of view.

If we assume that we have a global fibration, several cases could also occur:
\begin{itemize}
 \item the fibers could be isomorphic to quotients of $PSL(2,\C)$ by a
Kleinian group, i.e. to the orthonormal frame bundle of
a quotient of the hyperbolic 3-space by the Kleinian group. In some cases such a quotient is compact and
thus an analogue of Theorem \ref{theoremEuclide} holds. We hence recover in this case a quotient
space-time $\mathcal{X}$ which is a solution of the classical Einstein equations. One may wonder
however whether the geometry of the fiber may have a physical impact, e.g. at the quantum level.
\item the fibers are copies of $SO(1,3)$ or $Spin(3)$ or not compact quotients as previously.
In this case Theorem \ref{theoremEuclide} does not hold in general, unless some further hypotheses
are assumed. Equations (\ref{firstsystem}) are then the Einstein--Cartan system of equations
with sources (the stress-energy tensor and the angular momentum tensor) due to the auxiliary
fields $p$. The main question is to understand the dynamics of the fields $\varpi$ or $p$,
governed by Equations \ref{secondsystem} and, probably, to understand what kind of hypotheses
one should impose on these fields.
\end{itemize}

\section{Gauge invariances}\label{sectioninvariance}
The action
\[
\mathcal{A}[\Gamma] = \int_\Gamma\theta
 = \int_\mathcal{P}\varphi^*\theta
\]
and the constraints (\ref{constraints4}) are invariant by the action of several gauge groups:
\begin{itemize}
 \item they are invariant \emph{off-shell} by orientation preserving
diffeomorphisms or by reparametrizations: if
$\phi:\mathcal{P}\longrightarrow \mathcal{P}$ is an
orientation preserving diffeomorphism, then
$\int_\mathcal{P}\varphi^*\theta = \int_\mathcal{P}(\varphi\circ \phi)^*\theta$;
\item they are invariant \emph{on-shell} by gauge transformations with structure gauge group
$\mathfrak{G}$: assume that $\varphi^*\eta = (\alpha,\omega)$ satisfies the two last
HVDW equations (\ref{decompostionOmegaAfteralpha}) and (\ref{decompostionOmegaAfteromega}),
then, by Lemma \ref{spontaneousfibration}, $\mathcal{P}$ looks everywhere locally
like a principal bundle over a 4-dimensional manifold $\mathcal{X}$ with structure
group $\mathfrak{G}$. In particular we can find local coordinates $(x,g)$ in which $(\alpha,\omega)$ reads
$\alpha = g^{-1}e$ and $\omega = g^{-1}dg + g^{-1}Ag$, where $(e,A)$ is a $\mathfrak{p}$-valued
1-form which depends only on $x$. The gauge group is then described locally as the set
of maps $\gamma:\mathcal{P}\longrightarrow \mathfrak{G}$ of the form $\gamma(x,g) = g^{-1}f(x)g$,
where $f$ is a map from $\mathcal{X}$ to $\mathfrak{G}$ and any such map $\gamma$ acts on
$(\alpha,\omega)$ by 
\[
 (\alpha,\omega) \longmapsto (\tilde{\alpha},\tilde{\omega}) =
(\gamma^{-1}\alpha, \gamma^{-1}d\gamma + \gamma^{-1}\omega\gamma)
= (0,\gamma^{-1}d\gamma) + \hbox{Ad}_{\gamma^{-1}}(\alpha,\omega)
\]
and on $\varpi$ by $\varpi \longmapsto \tilde{\varpi} = \hbox{Ad}_{\gamma}^*\varpi$.

Then $\varphi^*(d\eta+\frac{1}{2}[\eta\wedge \eta])$ is changed in
$\tilde{\varphi}^*(d\eta+\frac{1}{2}[\eta\wedge \eta]) =
\hbox{Ad}_{\gamma^{-1}}\left(\varphi^*(d\eta+\frac{1}{2}[\eta\wedge \eta])\right)$.
Hence the Lagrangian density $\varphi^*\left(\psi\wedge (d\eta+\frac{1}{2}[\eta\wedge \eta])\right)$
is left unchanged.
Moreover a computation similar to the proof of (\ref{tetedep}) shows that the constraint (\ref{constraints4}),
which reads also $\varpi{_A}^{ab} = \kappa_A^{ab}$, is preserved by this transformation.
Note that $\tilde{\alpha} = g^{-1}\tilde{e}$ and $\tilde{\omega} = g^{-1}\tilde{A}g +g^{-1}dg$,
where $\tilde{e} = f^{-1}e$ and $\tilde{A} = f^{-1}Af + f^{-1}df$ and thus
$d\tilde{e} + \tilde{A}\wedge \tilde{e} = f^{-1}(de+A\wedge e)$ and
$d\tilde{A}+\tilde{A}\wedge \tilde{A} = f^{-1}(dA+A\wedge A)f$. 

\item Lastly  we can write the action density as
\[
 \psi\wedge (d\eta+\frac{1}{2}[\eta\wedge \eta])
= -\left( d\psi - \hbox{ad}_\eta^*\wedge \psi\right)\wedge \eta + d(\psi\wedge \eta),
\]
which shows that, up to an exact term, the action is invariant \emph{off-shell} by transformations of the form
\[
 (\alpha,\omega,\varpi) \longmapsto (\alpha,\omega,\varpi + \chi),
\]
where $\chi$ is any $\mathfrak{p}^*$-valued 8-form with compact support which satisfies the condition
$ d\chi - \hbox{ad}_\eta^*\wedge \chi = 0$. If we moreover assume that $\chi\wedge \alpha^a\wedge \alpha^b = 0$,
$\forall a,b$, then the constraint (\ref{constraints4}) is also preserved.
\end{itemize}

\section{Annex}
 
\subsection{Lie algebras and their dual spaces}\label{annexeLie}
For the notations we refer to Section \ref{notations}. Moreover we denote by $(E^0,E^1,E^2,E^3)$
the basis of $\vec{\mathbb{M}}^*$ which is dual to $(E_0,E_1,E_2,E_3)$;
$(u^4,\cdots,u^9)$ is the basis of $\mathfrak{g}^*$ which is dual to $(u_4,\cdots,u_9)$ and
$(\mathfrak{l}^0,\cdots,\mathfrak{l}^9)$ is the basis of $\mathfrak{p}^*$ which is dual
$(\mathfrak{l}_0,\cdots,\mathfrak{l}_9)$.
The structure coefficients of $\mathfrak{p}$ in the basis
$(\mathfrak{l}_A)_{0\leq A\leq 9}$ are denoted by $c_{CB}^A$, so that $[\mathfrak{l}_B,\mathfrak{l}_C]
= c_{BC}^A\mathfrak{l}_A$, for $0\leq A,B,C\leq 9$.
In the subcase where $A,B,C = i,j,k$ run from 4 to 9, we recover
the structure coefficients of $\mathfrak{g}$ in the basis $(u_4,u_5,u_6,u_7,u_8,u_9)$,
i.e. such that $[u_j,u_k]= c^i_{jk}u_i$.

\subsubsection{Tensorial notations for $\mathfrak{g}$ and $\mathfrak{g}^*$}
Consider $\vec{\mathbb{M}}\otimes \vec{\mathbb{M}}$,
$\vec{\mathbb{M}}^*\otimes \vec{\mathbb{M}}^*$, $\vec{\mathbb{M}}\otimes \vec{\mathbb{M}}^*$
and $\vec{\mathbb{M}}^*\otimes \vec{\mathbb{M}}$ and their vector subspaces
$\vec{\mathbb{M}}\wedge \vec{\mathbb{M}}:= \{t^{ab}E_{ab}\in \vec{\mathbb{M}}\otimes\vec{\mathbb{M}};
\, t^{ab}+t^{ba} = 0\}$,
$\vec{\mathbb{M}}\wedge \vec{\mathbb{M}}^*:= \{ t{^a}_bE{_a}^b\in \vec{\mathbb{M}}\otimes\vec{\mathbb{M}}^*;
t{^a}_{b'}\textsf{h}^{b'b} + t{^b}_{a'}\textsf{h}^{a'a} = 0\}$,
$\vec{\mathbb{M}}^*\wedge \vec{\mathbb{M}}^*:= \{ t_{ab}E^{ab}\in \vec{\mathbb{M}}\otimes\vec{\mathbb{M}}^*;
t_{ab} + t_{ba} = 0\}$ and
$\vec{\mathbb{M}}^*\wedge \vec{\mathbb{M}}:= \{ t{_a}^bE{^a}_b\in \vec{\mathbb{M}}^*\otimes\vec{\mathbb{M}};
t{_a}^{b'}\textsf{h}_{b'b} + t{_b}^{a'}\textsf{h}_{a'a} = 0\}$,
where we write for short $E_{ab}:= E_a\otimes E_b$, $E{_a}^b:= E_a\otimes E^b$,
$E^{ab}:= E^a\otimes E^b$ and $E{^a}_b:= E^a\otimes E_b$.

To any $\xi\in \mathfrak{g}$ it corresponds a
unique tensor $\xi{^a}_bE{_a}^b \in \vec{\mathbb{M}}\wedge \vec{\mathbb{M}}^*$ such that
$\mathcal{R}(\xi)(E_b) = E_a\xi{^a}_{b}$ and conversely. Hence we
get the following vector spaces isomorphisms
\[
\begin{array}{cccl}
  \grave{\ell}: & \mathfrak{g} & \longrightarrow & \vec{\mathbb{M}}\wedge \vec{\mathbb{M}}^*\\
 & \xi & \longmapsto & \xi{^a}_bE{_a}^b
 \end{array}
\quad\hbox{and}\quad 
 \begin{array}{cccl}
  \bar{\ell}: & \mathfrak{g} & \longrightarrow & \vec{\mathbb{M}}\wedge \vec{\mathbb{M}}\\
& \xi & \longmapsto & \xi^{ab}E_{ab},
 \end{array}
\]
where $\xi^{ab} = \xi{^a}_{b'}\textsf{h}^{b'b}$.
We have the canonical identifications $\vec{\mathbb{M}}^*\wedge \vec{\mathbb{M}}\simeq (\vec{\mathbb{M}}\wedge \vec{\mathbb{M}}^*)^*$
and
$\vec{\mathbb{M}}^*\wedge \vec{\mathbb{M}}^*\simeq (\vec{\mathbb{M}}\wedge \vec{\mathbb{M}})^*$ 
by using respectively the duality pairings
\[
(\lambda{_{a'}}^{b'}E{^{a'}}_{b'} ,\xi{^a}_bE{_a}^b) \longmapsto \frac{1}{2}\lambda{_a}^b\xi{^a}_b
\quad \hbox{and}\quad
(\lambda_{a'b'}E^{a'b'},\xi^{ab}E_{ab}) \longmapsto \frac{1}{2}\lambda_{ab}\xi^{ab}.
\]
Through these identifications, the adjoints of $\grave{\ell}$ and $\bar{\ell}$ provides us with isomorphisms
$\grave{\ell}^*: \vec{\mathbb{M}}^*\wedge \vec{\mathbb{M}}\longrightarrow \mathfrak{g}^*$ and
$\bar{\ell}^*: \vec{\mathbb{M}}^*\wedge \vec{\mathbb{M}}^* \longrightarrow  \mathfrak{g}^*$.
We define $u^i_{ab}$ and $u^{ib}_a = u^i_{ab'}\textsf{h}^{b'b}$ by
$(\bar{\ell}^*)^{-1}(u^i) = u^i_{ab}E^{ab}$ and $(\grave{\ell}^*)^{-1}(u^i) = u^{ib}_aE{^a}_b$,
$\forall i = 4,\cdots,9$. We then have
\begin{equation}\label{a86}
 \frac{1}{2}u^i_{ab}u_j^{ab} = \frac{1}{2}u^{ib}_au_{jb}^a = u^i(u_j) = \delta^i_j
\end{equation}
and
\begin{equation}\label{a87}
 u^i_{ab}u_i^{a'b'} = \frac{1}{2}\delta^{a'b'}_{ab}:= \frac{1}{2}
(\delta^{a'}_a\delta^{b'}_b - \delta^{a'}_b\delta^{b'}_a),
\end{equation}
from which we also deduce
\begin{equation}\label{a88}
 u^{ib}_au^{a'}_{ib'} =  \frac{1}{2}
(\delta^{a'}_a\delta^b_{b'} - \textsf{h}^{a'b}\textsf{h}_{ab'}).
\end{equation}

\subsubsection{Tensorial notations for $\mathfrak{p}$}
We can extend the previous isomorphism $\grave{\ell}$ to
\[
 \begin{array}{cccl}
  \grave{\ell}: &\mathfrak{p} & \longrightarrow & (\vec{\mathbb{M}}\wedge \vec{\mathbb{M}}^*)\oplus \vec{\mathbb{M}}\\
 & \xi & \longmapsto & (\xi{^a}_bE{_a}^b,\xi^aE_a)
 \end{array}
\]
where, denoting by $O$ the origin of $\mathbb{M}$,
$\mathcal{R}(\xi)(O+x^aE_a) = O+ x^b\xi{^a}_{b}E_a + \xi^aE_a$, $\forall \xi\in \mathfrak{p}$,
$\forall O+x^bE_b\in \mathbb{M}$.
We then have
\[
 \grave{\ell}([\xi,\zeta]) = \left((\xi{^{a}}_{c}\zeta{^{c}}_{b} - \zeta{^{a}}_{c}\xi{^{c}}_{b})E{_a}^b,\,
(\xi{^{a}}_{b}\zeta{^{b}} - \zeta{^{a}}_{b}\xi{^{b}})E_a\right)
\]
As for $\mathfrak{g}^*$ we also get the following vector spaces isomorphism
\[
\begin{array}{cccl}
 (\grave{\ell}^*)^{-1}: & \mathfrak{p}^* & \longrightarrow & (\vec{\mathbb{M}}^*\wedge \vec{\mathbb{M}})\oplus \vec{\mathbb{M}}^*\\
& \lambda  & \longmapsto & (\lambda{_a}^bE{^a}_b,\lambda_aE^a)
\end{array}
\]
with the duality pairing
$\left((\vec{\mathbb{M}}^*\wedge \vec{\mathbb{M}})\oplus \vec{\mathbb{M}}^*\right)\times
\left((\vec{\mathbb{M}}\wedge \vec{\mathbb{M}}^*)\oplus\vec{\mathbb{M}}\right) \longrightarrow \R$,
\[
  \left((\lambda{_a}^bE{^a}_b,\lambda_aE^a),(\xi{^a}_bE{_a}^b,\xi^aE_a)\right)
\longmapsto \frac{1}{2}\lambda{_a}^b\xi{^a}_b + \lambda_a\xi^a.
\]

\subsubsection{Adjoint and coadjoint action of $\mathfrak{G}$}
The standard representation $\mathcal{R}$ of $\mathfrak{G}$ induces the map
$\mathfrak{G}\longrightarrow \vec{\mathbb{M}}\wedge \vec{\mathbb{M}}^*$, $g\longmapsto g{^a}_bE{_a}^b$.
The restriction to $\mathfrak{G}$ of the adjoint representation of $\mathfrak{P}$ on $\mathfrak{p}$
reads
\[
\forall \xi\in \mathfrak{p},\quad
 \hbox{Ad}_g(\xi{^a}_bE{_a}^b,\xi^aE{_a}) = \left((g{^a}_{a'}\xi{^{a'}}_{b'}(g^{-1}){^{b'}}_b)E{_a}^b,\,g{^a}_{a'} \xi^{a'}E_a\right).
\]
The coadjoint action of $\mathfrak{G}$ on $\mathfrak{p}^*$ is defined by:
$\forall g\in \mathfrak{G}$, $\forall \lambda\in \mathfrak{p}^*$, $\hbox{Ad}_g^*\lambda$
is the vector in $\mathfrak{p}^*$
such that:
\[
\forall \xi\in \mathfrak{p},\quad
(\hbox{Ad}_g^*\lambda)(\xi):= \lambda(\hbox{Ad}_g\xi).
\]
In our setting this reads
\[
 \begin{array}{ccl}
  (\hbox{Ad}_g^*\lambda)(\xi) & = & \frac{1}{2}\lambda{_a}^b
\left(g{^a}_{a'}\xi{^{a'}}_{b'}(g^{-1}){^{b'}}_b\right) +
 \lambda_a\left(g{^a}_{a'}\right) \xi^{a'}\\
& = & \frac{1}{2}\left(g{^{a'}}_{a}\lambda{_{a'}}^{b'}(g^{-1}){^{b}}_{b'}\right)\xi{^{a}}_{b}
+ \left(g{^{a'}}_{a}\lambda_{a'}\right)\xi^{a}.
 \end{array}
\]
Hence
\begin{equation}\label{a89}
 (\grave{\ell}^*)^{-1}(\hbox{Ad}_g^*\lambda) = \left((g{^{a'}}_{a}\lambda{_{a'}}^{b'}(g^{-1}){^{b}}_{b'})E{^a}_b,
\, g{^{a'}}_{a}\lambda_{a'}E^a\right).
\end{equation}

\subsubsection{Coadjoint action of $\mathfrak{p}$}
The coadjoint action of $\mathfrak{p}$ on $\mathfrak{p}^*$ is defined by:
$\forall \xi\in \mathfrak{p}$, $\forall \lambda\in \mathfrak{p}^*$, $\hbox{ad}_\xi^*\lambda$
is the vector in $\mathfrak{p}^*$
such that:
\[
\forall \zeta\in \mathfrak{p},\quad
(\hbox{ad}_\xi^*\lambda)(\zeta):= \lambda(\hbox{ad}_\xi\zeta) =  \lambda([\xi,\zeta]).
\]
This gives us:
\[
 \begin{array}{ccl}
  (\hbox{ad}_\xi^*\lambda)(\zeta) & = & \frac{1}{2}\lambda{_a}^b
\left(\xi{^{a}}_{c}\zeta{^{c}}_{b} - \zeta{^{a}}_{c}\xi{^{c}}_{b} \right)
+ \lambda_a\left( \xi{^{a}}_{b}\zeta{^{b}} - \zeta{^{a}}_{b}\xi{^{b}}\right)\\
& = & \frac{1}{2}\left(\xi{^{c}}_{a} \lambda{_{c}}^{b} - \lambda{_a}^c\xi{^{b}}_{c}
- 2\lambda_a\xi{^{b}}\right)\zeta{^{a}}_{b} + \left(\xi{^{a}}_{b}\lambda_a\right)\zeta{^{b}}
 \end{array}
\]
Hence
\begin{equation}\label{deficoadjoint}
 (\grave{\ell}^*)^{-1}(\hbox{ad}_\xi^*\lambda) = \left((\xi{^{c}}_{a} \lambda{_{c}}^{b}
- \lambda{_a}^c\xi{^{b}}_{c} - 2\lambda_a\xi{^{b}})E{^a}_b,\,\xi{^{a}}_{b}\lambda_aE^b\right).
\end{equation}
An alternative representation  uses the basis $(\mathfrak{l}_A)_A$ of $\mathfrak{p}$ and the dual basis
$(\mathfrak{l}^A)_A$ of $\mathfrak{p}^*$: decompose $\lambda=\lambda_A\mathfrak{l}^A$,
$\xi = \mathfrak{l}_A\xi^A$ and $\zeta = \mathfrak{l}_A\zeta^A$, then
$[\xi,\zeta] = \mathfrak{l}_Ac_{BC}^A\xi^B\zeta^C$, we find that
$(\hbox{ad}_\xi^*\lambda)(\xi) = \lambda_A(c_{BC}^A\xi^B\zeta^C) = 
(\lambda_Bc_{CA}^B\xi^C)\zeta^A$ hence
\[
 \hbox{ad}_\xi^*\lambda = (\lambda_Bc_{CA}^B\xi^C)\mathfrak{l}^A.
\]
We can extend this action to $\mathfrak{p}$-valued and $\mathfrak{p}^*$-valued exterior forms.
If $\xi$ is a $\mathfrak{p}$-valued form and $\lambda$ is a $\mathfrak{p}^*$-valued form,
we define
\begin{equation}\label{defiadxi*}
 \hbox{ad}_\xi^*\wedge \lambda:= c_{CA}^B(\xi^C\wedge \lambda_B)\mathfrak{l}^A.
\end{equation}
\begin{lemm}\label{lemmaAdad}
 Let $g\in \mathfrak{G}$, $\xi\in \mathfrak{p}$ and $\lambda\in \mathfrak{p}^*$.
 Then
\begin{equation}\label{formulaAdad}
 \hbox{Ad}_{g^{-1}}^*\left(\hbox{ad}_{(\hbox{Ad}_{g^{-1}}\xi)}^*\lambda\right) = 
\hbox{ad}_\xi^*(\hbox{Ad}_{g^{-1}}^*\lambda).
\end{equation}
\end{lemm}
\emph{Proof} --- Take any $\zeta\in \mathfrak{p}$ and start from the identity
$\hbox{Ad}_{g^{-1}}([\xi,\zeta]) = \left[\hbox{Ad}_{g^{-1}}\xi,\hbox{Ad}_{g^{-1}}\zeta\right]$,
which implies
\[
 \lambda\left(\hbox{Ad}_{g^{-1}}[\xi,\zeta]\right) =
\lambda\left(\hbox{ad}_{(\hbox{Ad}_{g^{-1}}\xi)} (\hbox{Ad}_{g^{-1}}\zeta)\right).
\]
The l.h.s. of this identity is equal to
$(\hbox{Ad}_{g^{-1}}^*\lambda)([\xi,\zeta])
= \left(\hbox{ad}_\xi^*\left(\hbox{Ad}_{g^{-1}}^*\lambda\right)\right)\left(\zeta\right)$
and its r.h.s. is equal to
$\left(\hbox{ad}_{(\hbox{Ad}_{g^{-1}}\xi)}^*\lambda\right)\left(\hbox{Ad}_{g^{-1}}\zeta\right)
= \left(\hbox{Ad}_{g^{-1}}^*\left(\hbox{ad}_{(\hbox{Ad}_{g^{-1}}\xi)}^*\lambda\right)\right)(\zeta)$.
Hence (\ref{formulaAdad}) follows.\hfill $\square$


\subsection{Exterior differential calculus}
\begin{lemm}\label{lemma0}
The following relations holds
\[
 \alpha^a\wedge \alpha_{a'}^{(4)} = \delta_{a'}^a\alpha^{(4)},\quad \alpha^a\wedge \alpha^b\wedge \alpha_{a'b'}^{(4)}
 = \delta^{ab}_{a'b'}\alpha^{(4)}
\]
and
\[
 \omega^i\wedge \omega_{i'}^{(6)} = \delta^i_{i'}\omega^{(6)},\quad \omega^i\wedge \omega^j\wedge \omega^{(6)}_{i'j'}
 = \delta^{ij}_{i'j'}\omega^{(6)}.
\]
where $\delta^{ab}_{a'b'}:= \delta^a_{a'}\delta^b_{b'} -  \delta^a_{b'}\delta^b_{a'}$
and $\delta^{ij}_{i'j'}:= \delta^i_{i'}\delta^j_{j'} -  \delta^i_{j'}\delta^j_{i'}$.
\end{lemm}
The proof is left to the Reader.
\begin{lemm}\label{lemma1}
Let $e^{(4)}:= e^0\wedge e^1\wedge e^2\wedge e^3$ and
$e_{ab}^{(2)}:= \frac{\partial}{\partial e^b}\iN 
 \frac{\partial}{\partial e^a}\iN e^{(4)}$. Then
\[
 e_{ab}^{(2)} = \frac{1}{2}\epsilon_{abcd}e^c\wedge e^d
\] 
\end{lemm}
\emph{Proof} --- We start from
$e^{(4)} = \frac{1}{4!}\epsilon_{a'b'c'd'}e^{a'}\wedge e^{b'}\wedge e^{c'}\wedge e^{d'}$.
We then compute $e_a^{(3)} :=  \frac{\partial}{\partial e^a}\iN e^{(4)}$:
\[
 \begin{array}{c}
   \displaystyle e_a^{(3)} = 1/4!\left[\epsilon_{ab'c'd'}e^{b'}\wedge e^{c'}\wedge e^{d'}
 - \epsilon_{a'ac'd'}e^{a'}\wedge e^{c'}\wedge e^{d'} + \epsilon_{a'b'ad'}e^{a'}\wedge e^{b'}\wedge e^{d'}
 - \epsilon_{a'b'c'a}e^{a'}\wedge e^{b'}\wedge e^{c'}\right]\\
  \displaystyle = 1/4!\left[\epsilon_{ab'c'd'}e^{b'}\wedge e^{c'}\wedge e^{d'}
 + \epsilon_{aa'c'd'}e^{a'}\wedge e^{c'}\wedge e^{d'}+ \epsilon_{aa'b'd'}e^{a'}\wedge e^{b'}\wedge e^{d'}
 + \epsilon_{aa'b'c'}e^{a'}\wedge e^{b'}\wedge e^{c'}\right]\\
 \displaystyle = 1/3!\, \epsilon_{ab'c'd'}e^{b'}\wedge e^{c'}\wedge e^{d'}.
 \end{array}
\]
By performing a similar computation for $e_{ab}^{(2)}$ we obtain the result. \hfill $\square$\\
\emph{Corollary} --- We deduce from the lemma that 
$e_{cd'}^{(2)}\textsf{h}^{d'd} = \frac{1}{2}\epsilon{_{abc}^{}}^{d'}e^a\wedge e^b$, hence
\[
 \textsf{h}^{dd'}e_{cd}^{(2)}\wedge \Omega{^c_{}}_{d'} = \frac{1}{2}\epsilon{_{abc}^{}}^{d'}e^a\wedge e^b\wedge \Omega{^c_{}}_{d'}.
\]
\begin{lemm}\label{lemma2}
Let $\gamma:= g^{-1}dg$ be the Maurer--Cartan form on the group $\mathfrak{G}$,
$(\gamma^i)_{1\leq i\leq 6}$ the components of $\gamma$
in a basis $(\mathfrak{t}_1,\cdots, \mathfrak{t}_6)$ of $\mathfrak{g}$,
$\gamma^{(6)}:= \gamma^1\wedge \cdots \wedge \gamma^6$,
$\gamma_i^{(5)}:= \frac{\partial}{\partial \gamma^i}\iN \gamma^{(6)}$,
$\gamma_{ij}^{(4)}:= \frac{\partial}{\partial \gamma^j}\iN \frac{\partial}{\partial \gamma^i}\iN \gamma^{(6)}$.
Lastly let $c^i_{jk}$ be the structure constants of $\mathfrak{g}$ in the basis
$(\mathfrak{t}_1,\cdots, \mathfrak{t}_6)$.
Then
\begin{equation}\label{93}
 d\gamma^i + \frac{1}{2}c^i_{jk}\gamma^j\wedge \gamma^k =  0,
\end{equation}
\begin{equation}\label{dgamma6}
 d\gamma^{(6)} = 0,
\end{equation}
\begin{equation}\label{dgamma5}
 d\gamma_i^{(5)} = 0
\end{equation}
\begin{equation}\label{dgamma4}
 d\gamma_{ij}^{(4)} + c^k_{ij}\gamma_k^{(5)} = 0.
\end{equation}
\end{lemm}
\emph{Proof} --- Relation (\ref{dgamma6}) is simply due to the fact that $\gamma^{(6)}$ has a
maximal degree. Relation (\ref{dgamma5}) follows from (\ref{93}) and the fact that $\mathfrak{g}$ is unimodular:
\[
 \begin{array}{ccl}
   d\gamma_i^{(5)} & = & d\gamma^j\wedge \gamma_{ij}^{(4)} =
- \frac{1}{2}c^j_{kl}\gamma^k\wedge \gamma^l\wedge \gamma_{ij}^{(4)}\\
    & = & - \frac{1}{2}c^j_{ij}\gamma^{(6)} + \frac{1}{2}c^j_{ji}\gamma^{(6)} = - c^j_{ij}\gamma^{(6)} = 0.
 \end{array}
\]
The reasoning is similar for (\ref{dgamma4}):
\[
 \begin{array}{ccl}
  d\gamma_{ij}^{(4)} & = & d\gamma^k\wedge\gamma_{ijk} = - \frac{1}{2}c^k_{lm}\gamma^l\wedge \gamma^m\wedge \gamma_{ijk}^{(3)}\\
  & = & - \frac{1}{2}c^k_{lm}\left[\delta^{lm}_{ij}\gamma_k^{(5)} + \delta^{lm}_{jk}\gamma_i^{(5)} + \delta^{lm}_{ki}\gamma_j^{(5)} \right]\\
   & = & - c^k_{ij}\gamma_k^{(5)} - c^k_{jk}\gamma_i^{(5)} - c^k_{ki}\gamma_j^{(5)} = - c^k_{ij}\gamma_k^{(5)}.
 \end{array}
\]
\hfill $\square$
\begin{lemm}\label{lemma3}
 Let $g$ be smooth map with values in $\mathfrak{G}$ and let $\varpi$ be an exterior differential form with
coefficients in $\mathfrak{p}^*$. Then
\begin{equation}\label{formulaLemma3}
 d\left(\hbox{Ad}_{g^{-1}}^*\varpi\right) =
\hbox{Ad}_{g^{-1}}^*\left(d\varpi - \hbox{ad}_{g^{-1}dg}^*\wedge \varpi\right).
\end{equation}
\end{lemm}
\emph{Proof} --- Assume that $\varpi$ is of degree $q$ and
consider any constant $\xi\in \mathfrak{p}$. We have 
\[
 d\left(\hbox{Ad}_{g^{-1}}^*\varpi\right)(\xi) =
d\left(\hbox{Ad}_{g^{-1}}^*\varpi(\xi)\right) =
d\left[\varpi\left(\hbox{Ad}_{g^{-1}}\xi\right)\right] =
(d\varpi) \left(\hbox{Ad}_{g^{-1}}\xi\right) + (-1)^q\varpi\wedge d\left(\hbox{Ad}_{g^{-1}}\xi\right).
\]
But since $d\left(\hbox{Ad}_{g^{-1}}\xi\right)
= -\hbox{ad}_{g^{-1}dg}\left(\hbox{Ad}_{g^{-1}}\xi\right)$ we deduce
\[
 \begin{array}{ccl}
  d\left(\hbox{Ad}_{g^{-1}}^*\varpi\right)(\xi) & = &
(d\varpi) \left(\hbox{Ad}_{g^{-1}}\xi\right) -
\left(\hbox{ad}_{g^{-1}dg}^*\wedge \varpi\right)\left(\hbox{Ad}_{g^{-1}}\xi\right)\\
& = & \left(d\varpi - \hbox{ad}_{g^{-1}dg}^*\wedge \varpi\right)\left(\hbox{Ad}_{g^{-1}}\xi\right)\\
& = & \left(\hbox{Ad}_{g^{-1}}^*\left(d\varpi - \hbox{ad}_{g^{-1}dg}^*\wedge \varpi\right)\right)(\xi).
 \end{array}
\]
Hence (\ref{formulaLemma3}) follows.
\hfill $\square$
\begin{coro}
 If $p:= \hbox{Ad}_{g^{-1}}^*\varpi$ and 
$(\alpha,\omega) = (0,g^{-1}dg) + \hbox{Ad}_{g^{-1}}H$, then
\begin{equation}\label{lastcoro}
 dp - \hbox{ad}_H^*\wedge p = \hbox{Ad}_{g^{-1}}^*\left(d\varpi
- \hbox{ad}_{(\alpha,\omega)}\wedge \varpi\right).
\end{equation}
\end{coro}
\emph{Proof} -- From (\ref{formulaAdad}) we deduce 
\[
 \hbox{ad}_H^*\wedge p = \hbox{ad}_H^*\wedge \left(\hbox{Ad}_{g^{-1}}^*\varpi\right)
= \hbox{Ad}_{g^{-1}}^*\left(\hbox{ad}_{(\hbox{Ad}_{g^{-1}}H)}\wedge \varpi\right),
\]
which, together with (\ref{formulaLemma3}), implies (\ref{lastcoro}).
\hfill $\square$



\end{document}